\documentclass[12pt]{scrartcl}
\usepackage{a4}
\usepackage{amsthm}
\usepackage{amsmath}
\usepackage{amssymb}
\usepackage{amsfonts}
\usepackage{mathrsfs}
\usepackage{latexsym}
\usepackage{color}
\usepackage{bbm}
\definecolor{Myblue}{rgb}{0,0,0.6}
\usepackage[a4paper,colorlinks,citecolor=Myblue,linkcolor=Myblue,urlcolor=Myblue,pdfpagemode=None]{hyperref}
\usepackage[square,numbers,sort&compress]{natbib}
\usepackage[all,cmtip]{xy}
\usepackage{tikz}

  \vfuzz \hfuzz
\def\nicecolourscheme{\shadedraw[top color=blue!5, bottom color=blue!35, draw=blue!50!black, dashed]}

\newcommand{\E}{\text{e}}
\newcommand{\I}{\text{i}}
\newcommand{\C}{\mathbbm{C}}

\newcommand{\N}{\mathbbm{N}}

\newcommand{\Z}{\mathbbm{Z}}
\def\1{\ifmmode\mathrm{1\!l}\else\mbox{\(\mathrm{1\!l}\)}\fi}
\newcommand{\one}{\mathbbm{1}}
\newcommand{\be}{\begin{equation}}
\newcommand{\ee}{\end{equation}}
\newcommand{\bes}{\begin{equation*}}
\newcommand{\ees}{\end{equation*}}

\newcommand{\sVir}{\mathsf{sVir}}
\newcommand{\MF}{\text{MF}_{\text{bi}}}
\newcommand{\MFR}{\text{MF}^\text{R}_{\text{bi}}}
\newcommand{\DG}{\text{DG}_{\text{bi}}}
\newcommand{\DGR}{\text{DG}^\text{R}_{\text{bi}}}
\newcommand{\tc }{\otimes_\C}
\newcommand{\tr}{\otimes_R}
\newcommand{\id}{\text{id}}

\newcommand\nxt{\noindent\raisebox{.08em}{\rule{.44em}{.44em}}\hspace{.4em}}
\newcommand\arxiv[2]      {\href{http://arXiv.org/abs/#1}{#2}}
\newcommand\doi[2]        {\href{http://dx.doi.org/#1}{#2}}
\newcommand\httpurl[2]    {\href{http://#1}{#2}}

\allowdisplaybreaks

\deffootnote[1em]{1em}{1em}
{\textsuperscript{\thefootnotemark}}

\theoremstyle{definition}
\newtheorem{definition}{Definition}

\newtheorem{theorem}[definition]{Theorem}
\newtheorem{lemma}[definition]{Lemma}
\newtheorem*{remark}{Remark}

\newtheorem{conjecture}[definition]{Conjecture}

\numberwithin{equation}{section}
\numberwithin{definition}{section}
\numberwithin{figure}{section}

\begin{document}

\title{On the monoidal structure of matrix bi-factorisations}
\author{Nils Carqueville\thanks{Address from October 1$^{\text{st}}$ 2009 onwards: Arnold Sommerfeld Center for Theoretical Physics, LMU M\"unchen, Theresienstra\ss e~37, D-80333 M\"unchen \& Excellence Cluster Universe, Boltzmannstra\ss e~2, D-85748 Garching}, \quad Ingo Runkel\thanks{Address from October 1$^{\text{st}}$ 2009 onwards: Department Mathematik, Universit\"{a}t Hamburg, Bundesstra\ss e 55, D-20146 Hamburg}
\\[0.5cm]
  {\normalsize\slshape King's College London, Department of Mathematics,}\\[-0.1cm]
  {\normalsize\slshape Strand, London WC2R\,2LS, UK}\\[-0.1cm]
  \normalsize{\tt \href{mailto:nils.carqueville@kcl.ac.uk}{nils.carqueville@kcl.ac.uk}}, \quad
  \normalsize{\tt \href{mailto:ingo.runkel.ac.uk}{ingo.runkel@kcl.ac.uk}}}
\date{}
\maketitle

\vspace{-11.5cm}
\hfill {\scriptsize KCL-MTH-09-11}

\vspace{12cm}

\begin{abstract}
We investigate tensor products of matrix factorisations. This is most naturally done by formulating matrix factorisations in terms of bimodules instead of modules. If the underlying ring is $\C[x_1,\ldots,x_N]$ we show that bimodule matrix factorisations form a monoidal category.

This monoidal category has a physical interpretation in terms of defect lines in a two-dimensional Landau-Ginzburg model. There is a dual description via conformal field theory,  which in the special case of $W=x^d$ is an $\mathcal N=2$ minimal model, and which also gives rise to a monoidal category describing defect lines. We carry out a comparison of these two categories in certain subsectors by explicitly computing 6j-symbols.
\end{abstract}

\newpage

\tableofcontents

\section{Introduction and summary}\label{introduction}

Matrix factorisations first appeared in mathematics in the study of maximal Cohen-Macaulay modules~\cite{eisen1980,buchweitzMCM}. Later it was realised in the context of string theory that they also describe boundary conditions in two-dimensional $\mathcal N=2$ supersymmetric Landau-Ginzburg models~\cite{KontsevichU,kl0210,bhls0305,l0312}. Matrix factorisations form a category, and in the relation to Landau-Ginzburg models the objects of this category are interpreted as boundary conditions and the morphisms as boundary fields. The same structure appears in the study of two-dimensional conformal field theory. There, a boundary condition can be defined as a solution to a set of consistency conditions, the factorisation or sewing constraints~\cite{Cardy:1991tv,Lewellen:1991tb,Kong2006b,Fjelstad:2006aw}, and again one obtains a category whose objects and morphisms correspond to boundary conditions and boundary fields, respectively.

Mathematically these two descriptions are very different, one is the category of matrix factorisations over a polynomial ring, the other is constructed from the representations of a vertex operator algebra. However, physically one expects that the infrared fixed point of the $\mathcal N=2$ Landau-Ginzburg model is an $\mathcal N=2$ supersymmetric conformal field theory, and that the supersymmetry preserving boundary conditions in these two theories should be related. This physical reasoning has been tested by matching subsectors of the two categories of boundary conditions, see e.\,g.~\cite{add0401,h0401,bg0503,err0508}.

Apart from boundary conditions, there is another natural structure for two-dimensional field theories one can study, namely defect lines. A defect line is a line of inhomogeneity on the two-dimensional worldsheet, and the possible ``defect conditions'' again form a category. This is not surprising, because via the folding trick \cite{Wong:1994pa} one can think of a defect line in a given theory as a boundary for the corresponding folded theory, and so the same mathematical framework applies. However, there is one crucial difference between boundaries and defects, and this is that several defect lines can meet in junction points. As we will outline below, defect junctions can be encoded by endowing the category of defects with a monoidal structure. 

On the side of matrix factorisations, the relevant tensor product is known from \cite{yoshinoTP,add0401,Khovanov:2004,br0707.0922}, but the full monoidal structure, including the associator and unit isomorphisms, has not yet been coherently given. We close this gap in the case of polynomial rings $R=\C[x_1,\ldots,x_N]$ and prove that defects in any topologically B-twisted Landau-Ginzburg model indeed form a monoidal category. Some of the necessary ingredients are already contained in \cite{Khovanov:2004,br0707.0922};  we formulate them in the language of bimodule matrix factorisations -- or \textit{matrix bi-factorisations} for short -- and show that they satisfy the necessary coherence conditions for a monoidal category (theorems~\ref{thm:MF} and~\ref{thm:MFBmany}). We also prove a corresponding statement for graded matrix bi-factorisations (theorems~\ref{thm:MFR} and~\ref{thm:MFBmany}).

The infrared fixed point of the $\mathcal N=2$ Landau-Ginzburg model with superpotential $W=x^d$ is the A-type $\mathcal N=2$ super-Virasoro minimal model of central charge $c = 3-\frac{6}{d}$~\cite{kms1989,m1989,vw1989,howewest}, and one can now try to compare the corresponding monoidal categories of defect conditions. Before describing the monoidal structure and the comparison calculations, let us motivate in more detail how the tensor product arises on the category of defect conditions.

\subsubsection*{Defects in two-dimensional quantum field theory}

A defect line can either be an oriented closed loop, or an oriented interval joining two field insertion points. Each field insertion can be the start and end point of an arbitrary finite number of defect lines, resulting in $n$-valent defect junctions, $n=0,1,2,\dots$. A typical piece of a worldsheet might look as follows:
$$
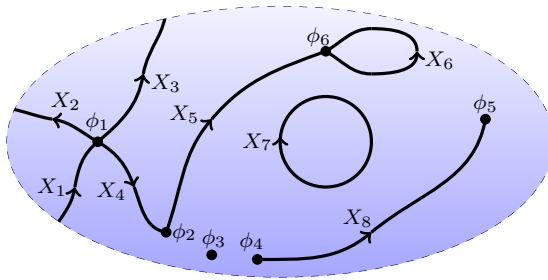
\begin{tikzpicture}[scale=0.6,baseline]
\clip (0,0) ellipse (6cm and 3cm);
\nicecolourscheme (0,0) ellipse (6cm and 3cm);
\draw (-5,-1) node {{\scriptsize $X_1$}};
\draw (-4.7,0.9) node {{\scriptsize $X_2$}};
\draw (-2.5,1.3) node {{\scriptsize $X_3$}};
\draw (-3.7,-1.1) node {{\scriptsize $X_4$}};
\draw (-2.1,0.6) node {{\scriptsize $X_5$}};
\draw (3.5,1.8) node {{\scriptsize $X_6$}};
\draw (-0.5,0) node {{\scriptsize $X_7$}};
\draw (1.7,-1.7) node {{\scriptsize $X_8$}};

\draw (-4,0.45) node {{\scriptsize $\phi_1$}};
\draw (-2.1,-2) node {{\scriptsize $\phi_2$}};
\draw (-1.45,-2.15) node {{\scriptsize $\phi_3$}};
\draw (-0.62,-2.2) node {{\scriptsize $\phi_4$}};
\draw (4.5,0.84) node {{\scriptsize $\phi_5$}};
\draw (0.85,2.32) node {{\scriptsize $\phi_6$}};

\draw[->, very thick, out=60, in=260] (-5,-2) to (-4.5,-1);
\draw[very thick, out=80, in=220] (-4.5,-1) to (-4,0);
\filldraw [black] (-4,0) circle (3pt);
\draw[->,very thick, out=150, in=350] (-4,0) to (-5,0.5);
\draw[very thick, out=170, in=350] (-5,0.5) to (-6,0.75);

\draw[->,very thick, out=-30, in=110] (-4,0) to (-3.2,-1);
\draw[very thick, out=-70, in=170] (-3.2,-1) to (-2.5,-2);
\filldraw [black] (-2.5,-2) circle (3pt);

\draw[->,very thick, out=70, in=230] (-2.5,-2) to (-1.5,0.5);
\draw[very thick, out=50, in=200] (-1.5,0.5) to (1,2);
\filldraw [black] (1,2) circle (3pt);

\draw[very thick, out=-30, in=180] (1,2) to (2,1.5);
\draw[->,very thick, out=0, in=270] (2,1.5) to (3,2);
\draw[very thick, out=90, in=0] (3,2) to (2,2.5);
\draw[very thick, out=180, in=30] (2,2.5) to (1,2);

\draw[very thick] (1,0) circle (1);
\draw[->,very thick, out=90, in=270] (0,0) to (0,0);

\filldraw [black] (-1.5,-2.5) circle (3pt);

\filldraw [black] (-0.5,-2.6) circle (3pt);
\draw[->,very thick, out=0, in=220] (-0.5,-2.6) to (2,-2);
\draw[very thick, out=40, in=260] (2,-2) to (4.5,0.5);
\filldraw [black] (4.5,0.5) circle (3pt);

\draw[->,very thick, out=40, in=260] (-4,0) to (-3,1.5);
\draw[very thick, out=80, in=270] (-3,1.5) to (-2.5,3);
\end{tikzpicture}
$$
Here we label the defect lines by defect conditions $X_i$ and the field insertions by fields $\phi_i$ from the corresponding space of $n$-valent junction fields. The quantum field theory assigns a value, the correlator, to each such worldsheet. These correlators are subject to consistency conditions arising from factorising a complicated worldsheet into simpler building blocks, see \cite{Runkel:2008gr,Calinetal}.

We are particularly interested in a special kind of defect condition, namely {\em topological defects}. These have the additional property that the path of the defect line can be deformed on the worldsheet without affecting the value of the correlator, as long as the defect line is not taken across field insertions or other defect lines. If we move two defect lines labelled by topological defect conditions $X$ and $Y$ arbitrarily close to each other, we obtain a new topological defect, the {\em fused defect}, and we write its defect condition as $X \otimes Y$:
\be
 \left\langle
\begin{tikzpicture}[scale=1,baseline]
\clip (0,0) ellipse (1.6cm and 1.2cm);
\draw[step=.5cm,gray,very thin] (-1.4,-1.4) grid (1.4,1.4);
\nicecolourscheme (0,0) ellipse (1.6cm and 1.2cm);
\draw (-0.55,-0.05) node {{\scriptsize $X$}};
\draw (0.08,-0.05) node {{\scriptsize $Y$}};
\draw[->,very thick, color=black] (-0.3,-1.5) -- (-0.3,0);
\draw[->,very thick, color=black] (0.3,-1.5) -- (0.3,0);
\draw[very thick, color=black] (-0.3,0) -- (-0.3,1.5);
\draw[very thick, color=black] (0.3,0) -- (0.3,1.5);
\end{tikzpicture}
\right\rangle
=
\left\langle
\begin{tikzpicture}[scale=1,baseline]
\clip (0,0) ellipse (1.6cm and 1.2cm);
\draw[step=.5cm,gray,very thin] (-1.4,-1.4) grid (1.4,1.4);
\nicecolourscheme (0,0) ellipse (1.6cm and 1.2cm);
\draw (-0.52,-0.05) node {{\scriptsize $X\otimes Y$}};
\draw[->,very thick, color=black] (0,-1.5) -- (0,0);
\draw[very thick, color=black] (0,0) -- (0,1.5);
\end{tikzpicture}
\right\rangle .
\ee
In these pictures we show a fraction of the worldsheet, and it is understood that the two worldsheets differ only in the way indicated.
The equality means that the correlators assigned to the two worldsheets coincide. We will denote the trivial defect -- or identity defect -- by ${\bf 1}$. It acts as the identity under fusion: ${\bf 1} \otimes X \cong X$ and $X \otimes {\bf 1} \cong X$.

Similarly, we are interested in {\em topological junction fields}, which are junction fields that can be moved on the worldsheet without affecting the amplitude. Not all junction fields have this property, and whether or not there are any at all depends on the defect conditions placed on the defect lines joining at a junction. Suppose we have a defect junction with $m$ defect lines $X_1,\dots,X_m$ pointing towards the junction and $n$ defect lines $Y_1,\dots,Y_n$ pointing away from it:
\be\label{eq:m->n-junction}
\begin{tikzpicture}[scale=1.5,baseline]
\clip (0,0) ellipse (1.6cm and 1.2cm);
\draw[step=.25cm,gray,very thin] (-1.4,-1.4) grid (1.4,1.4);
\nicecolourscheme (0,0) ellipse (1.6cm and 1.2cm);
\draw[very thick, color=black] (230:0.8) -- (0,0);
\draw (-0.71,-0.53) node {{\scriptsize $X_1$}};
\draw[->,very thick, color=black] (230:2) -- (230:0.8);
\draw[very thick, color=black] (265:0.8) -- (0,0);
\draw (-0.28,-0.79) node {{\scriptsize $X_2$}};
\draw[->,very thick, color=black] (265:2) -- (265:0.8);
\draw[dotted] (275:0.8) arc (275:317:0.8);
\draw[very thick, color=black] (340:0.8) -- (0,0);
\draw (0.73,-0.45) node {{\scriptsize $X_m$}};
\draw[->,very thick, color=black] (340:2) -- (340:0.8);

\draw[->,very thick, color=black] (0,0) -- (165:0.8);
\draw (-0.75,0) node {{\scriptsize $Y_1$}};
\draw[very thick, color=black] (165:0.8) -- (165:2);
\draw[->,very thick, color=black] (0,0) -- (130:0.8);
\draw (-0.55,0.4) node {{\scriptsize $Y_2$}};
\draw[very thick, color=black] (130:0.8) -- (130:2);
\draw[dotted] (120:0.8) arc (120:77:0.8);
\draw[->,very thick, color=black] (0,0) -- (50:0.8);
\draw (0.36,0.69) node {{\scriptsize $Y_n$}};
\draw[very thick, color=black] (50:0.8) -- (50:2);

\filldraw [black] (0,0) circle (1.5pt);
\end{tikzpicture}
\end{equation}
We denote the $\C$-vector space of topological junction fields by\footnote{%
Of course, one can have more general configurations of defect lines than those where the defect lines pointing towards and away from the junction are grouped together as in \eqref{eq:m->n-junction}. Because the defect lines have an orientation, there is a conjugation on the defect conditions $X \mapsto X^\vee$, corresponding to orientation reversal, and the space of junction fields in \eqref{eq:many-to-many-junction} is then isomorphic to $\text{Hom}_{\mathcal D}(X_1 \otimes \ldots \otimes X_m \otimes Y_n^\vee \otimes \ldots \otimes Y_1^\vee,{\bf 1})$. The topological junction fields for a more general arrangement of the orientations of the defect lines are obtained by placing the conjugations $(\,\cdot\,)^\vee$ accordingly.}
\be
 \label{eq:many-to-many-junction}
 \text{Hom}_{\mathcal D}(X_1 \otimes \ldots \otimes X_m , Y_1 \otimes \ldots \otimes Y_n) \, .
\ee
The defect category $\mathcal{D}$ has as objects the different defect conditions, and as morphisms the topological junction fields between them. The fusion of defect lines should endow $\mathcal{D}$ with the structure of a monoidal category, and indeed it does so in the theories we will consider.

Suppose that $\mathcal{D}$ has a semi-simple subsector with finitely many isomorphism classes of simple objects that is closed under fusion. Denote this subsector by $\mathcal{D}'$ and the simple objects by $D_i$, $i \in \mathcal{I}$ for some index set $\mathcal{I}$. The $D_i$ label elementary topological defects, which means that up to scalars the identity field $\id_{D_i}$ on the defect $D_i$ is the unique topological (2-valent) junction field,
$\text{Hom}_{\mathcal D}(D_i , D_i) = \C \cdot \id_{D_i}$. The fusion of two defects decomposes into a direct sum
\be
 D_i \otimes D_j \cong \bigoplus_{k \in \mathcal{I}} N_{ij}^{k} D_k  \, ,
\ee
for some non-negative integers $N_{ij}^{k}$ called fusion coefficients. To keep the notation at bay, let us assume that $N_{ij}^{k} \in \{0,1\}$. This is the case in the examples we will study below. For a triple $i,j,k \in \mathcal{I}$ such that $N_{ij}^{k} = 1$, pick a non-zero topological junction field
\be \label{eq:phi-ijk-choice}
 \begin{tikzpicture}[scale=1,baseline]
\clip (0,0) ellipse (1.6cm and 1.2cm);
\draw[step=.5cm,gray,very thin] (-1.4,-1.4) grid (1.4,1.4);
\nicecolourscheme (0,0) ellipse (1.6cm and 1.2cm);
\draw (0.31,0.14) node {{\scriptsize $\varphi_{ij}^k$}};
\draw (-0.8,-0.42) node {{\scriptsize $D_i$}};
\draw (0.77,-0.42) node {{\scriptsize $D_j$}};
\draw (-0.3,0.63) node {{\scriptsize $D_k$}};

\draw[very thick, color=black] (-0.5,-0.5) -- (0,0);
\draw[->,very thick, color=black] (-1.5,-1.5) -- (-0.5,-0.5);
\draw[->,very thick, color=black] (0,0) -- (-0,0.7);
\draw[very thick, color=black] (0,0.7) -- (0,1.5);
\draw[->,very thick, color=black] (1.5,-1.5) -- (0.5,-0.5);
\draw[very thick, color=black] (0.5,-0.5) -- (0,-0);

\filldraw [black] (0,0) circle (2.3pt);
\end{tikzpicture} \, , \quad
 \varphi_{ij}^k \in \text{Hom}_{\mathcal D}(D_i \otimes D_j, D_k) \, .
\ee
These can then be used to build junctions of higher valency. For example, one can obtain a basis of the space $\text{Hom}_{\mathcal D}(D_i \otimes D_j \otimes D_k, D_l)$ by first using the junction $\varphi_{ij}^p$, for all $p$, and then $\varphi_{pk}^l$:
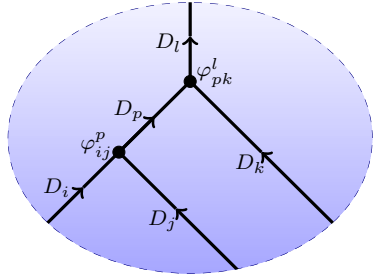
$$
 \begin{tikzpicture}[scale=1.5,baseline]
\clip (0,0) ellipse (1.6cm and 1.2cm);
\draw[step=.25cm,gray,very thin] (-1.4,-1.4) grid (1.4,1.4);
\nicecolourscheme (0,0) ellipse (1.6cm and 1.2cm);
\draw (-0.82,-0.07) node {{\scriptsize $\varphi_{ij}^p$}};
\draw (0.22,0.58) node {{\scriptsize $\varphi_{pk}^l$}};
\draw (-1.17,-0.44) node {{\scriptsize $D_i$}};
\draw (-0.24,-0.73) node {{\scriptsize $D_j$}};
\draw (0.53,-0.23) node {{\scriptsize $D_k$}};
\draw (-0.54,0.24) node {{\scriptsize $D_p$}};
\draw (-0.19,0.84) node {{\scriptsize $D_l$}};

\draw[->,very thick, color=black] (0,0.5) -- (0,0.9);
\draw[very thick, color=black] (0,0.9) -- (0,2);

\draw[->,very thick, color=black] (-0.625,-0.125) -- (-0.3125,0.1875);
\draw[very thick, color=black] (-0.375,0.125) -- (0,0.5);

\draw[->,very thick, color=black] (-1.5,-1) -- (-0.9375,-0.4375);
\draw[very thick, color=black] (-0.9375,-0.4375) -- (-0.375,0.125);

\draw[very thick, color=black] (0.625,-0.125) -- (0,0.5);
\draw[->,very thick, color=black] (1.5,-1) -- (0.625,-0.125);

\draw[very thick, color=black] (-0.125,-0.625) -- (-0.625,-0.125);
\draw[->,very thick, color=black] (0.5,-1.25) -- (-0.125,-0.625);

\filldraw [black] (0,0.5) circle (1.5pt);
\filldraw [black] (-0.625,-0.125) circle (1.5pt);
\end{tikzpicture}
$$
Similarly, one can build a basis by starting with $\varphi_{jk}^q$, for all $q$, and then use~$\varphi_{iq}^l$.  The two bases are related by a basis transformation, i.\,e.\ there have to be constants $F^{(ijk)l}_{pq} \in \C$ such that
\be
 \label{eq:defect-F-def1}
\left\langle
\begin{tikzpicture}[scale=1.5,baseline]
\clip (0,0) ellipse (1.6cm and 1.2cm);
\draw[step=.25cm,gray,very thin] (-1.4,-1.4) grid (1.4,1.4);
\nicecolourscheme (0,0) ellipse (1.6cm and 1.2cm);
\draw (0.83,-0.06) node {{\scriptsize $\varphi_{jk}^p$}};
\draw (0.22,0.58) node {{\scriptsize $\varphi_{ip}^l$}};
\draw (-0.84,-0.05) node {{\scriptsize $D_i$}};
\draw (-0.08,-0.58) node {{\scriptsize $D_j$}};
\draw (0.83,-0.57) node {{\scriptsize $D_k$}};
\draw (0.22,0.05) node {{\scriptsize $D_p$}};
\draw (-0.19,0.84) node {{\scriptsize $D_l$}};

\draw[->,very thick, color=black] (0,0.5) -- (0,0.9);
\draw[very thick, color=black] (0,0.9) -- (0,2);

\draw[->,very thick, color=black] (0.625,-0.125) -- (0.3125,0.1875);
\draw[very thick, color=black] (0.375,0.125) -- (0,0.5);

\draw[->,very thick, color=black] (1.5,-1) -- (0.9375,-0.4375);
\draw[very thick, color=black] (0.9375,-0.4375) -- (0.375,0.125);

\draw[very thick, color=black] (-0.625,-0.125) -- (0,0.5);
\draw[->,very thick, color=black] (-1.5,-1) -- (-0.625,-0.125);

\draw[very thick, color=black] (0.125,-0.625) -- (0.625,-0.125);
\draw[->,very thick, color=black] (-0.5,-1.25) -- (0.125,-0.625);

\filldraw [black] (0,0.5) circle (1.5pt);
\filldraw [black] (0.625,-0.125) circle (1.5pt);
\end{tikzpicture}
\right\rangle
=
\sum_{q \in \mathcal{I}} F^{(ijk)l}_{pq} \cdot 
\left\langle
\begin{tikzpicture}[scale=1.5,baseline]
\clip (0,0) ellipse (1.6cm and 1.2cm);
\draw[step=.25cm,gray,very thin] (-1.4,-1.4) grid (1.4,1.4);
\nicecolourscheme (0,0) ellipse (1.6cm and 1.2cm);
\draw (-0.82,-0.07) node {{\scriptsize $\varphi_{ij}^q$}};
\draw (0.22,0.58) node {{\scriptsize $\varphi_{qk}^l$}};
\draw (-1.17,-0.44) node {{\scriptsize $D_i$}};
\draw (-0.24,-0.73) node {{\scriptsize $D_j$}};
\draw (0.53,-0.23) node {{\scriptsize $D_k$}};
\draw (-0.54,0.24) node {{\scriptsize $D_q$}};
\draw (-0.19,0.84) node {{\scriptsize $D_l$}};

\draw[->,very thick, color=black] (0,0.5) -- (0,0.9);
\draw[very thick, color=black] (0,0.9) -- (0,2);

\draw[->,very thick, color=black] (-0.625,-0.125) -- (-0.3125,0.1875);
\draw[very thick, color=black] (-0.375,0.125) -- (0,0.5);

\draw[->,very thick, color=black] (-1.5,-1) -- (-0.9375,-0.4375);
\draw[very thick, color=black] (-0.9375,-0.4375) -- (-0.375,0.125);

\draw[very thick, color=black] (0.625,-0.125) -- (0,0.5);
\draw[->,very thick, color=black] (1.5,-1) -- (0.625,-0.125);

\draw[very thick, color=black] (-0.125,-0.625) -- (-0.625,-0.125);
\draw[->,very thick, color=black] (0.5,-1.25) -- (-0.125,-0.625);

\filldraw [black] (0,0.5) circle (1.5pt);
\filldraw [black] (-0.625,-0.125) circle (1.5pt);
\end{tikzpicture}
\right\rangle
\ee
holds for all correlators. The constants $F^{(ijk)l}_{pq}$ are the entries of the so-called {\em fusing matrices}, and they depend on the choice of 3-valent junction fields in~\eqref{eq:phi-ijk-choice}. When comparing fusing matrices of different realisations of the category $\mathcal{D}'$, one thus has to take care to compare basis-independent information.

The fusing matrices may just as well be extracted using the opposite orientation on the defect lines. To this end pick junction fields $\bar\varphi^{ij}_k \in \text{Hom}_{\mathcal D}(D_k,D_i \otimes D_j)$ such that
\be
\left\langle
\begin{tikzpicture}[scale=1.5,baseline]
\clip (0,0) ellipse (1.6cm and 1.2cm);
\draw[step=.25cm,gray,very thin] (-1.4,-1.4) grid (1.4,1.4);
\nicecolourscheme (0,0) ellipse (1.6cm and 1.2cm);
\draw (0.0,0.3) node {{\scriptsize $\varphi_{ij}^k$}};
\draw (0,-0.29) node {{\scriptsize $\bar\varphi_{k}^{ij}$}};

\draw (-0.72,-0.05) node {{\scriptsize $D_i$}};
\draw (0.73,-0.05) node {{\scriptsize $D_j$}};
\draw (-0.2,-0.93) node {{\scriptsize $D_k$}};
\draw (-0.2,0.81) node {{\scriptsize $D_k$}};

\draw[->,very thick, color=black] (0,0.5) -- (0,0.875);
\draw[very thick, color=black] (0,0.875) -- (0,1.9);

\draw[->,very thick, color=black] (0,-0.5) arc (270:180:0.5);
\draw[->,very thick, color=black] (0,-0.5) arc (270:360:0.5);
\draw[very thick, color=black] (-0.5,0) arc (180:90:0.5);
\draw[very thick, color=black] (0.5,0) arc (0:90:0.5);

\draw[->,very thick, color=black] (0,-1.9) -- (0,-0.875);
\draw[very thick, color=black] (0,-0.875) -- (0,-0.5);

\filldraw [black] (0,0.5) circle (1.5pt);
\filldraw [black] (0,-0.5)  circle (1.5pt);
\end{tikzpicture}
\right\rangle
=
\left\langle
\begin{tikzpicture}[scale=1.5,baseline]
\clip (0,0) ellipse (1.6cm and 1.2cm);
\draw[step=.5cm,gray,very thin] (-1.4,-1.4) grid (1.4,1.4);
\nicecolourscheme (0,0) ellipse (1.6cm and 1.2cm);
\draw (-0.25,-0.05) node {{\scriptsize $D_k$}};
\draw[->,very thick, color=black] (0,-1.5) -- (0,0);
\draw[very thick, color=black] (0,0) -- (0,1.5);
\end{tikzpicture}
\right\rangle \, .
\ee
The $\bar\varphi_{k}^{ij}$ exist by semi-simplicity. Because the space $\text{Hom}_{\mathcal D}(D_k,D_i \otimes D_j)$ is one-dimensional (recall that $N_{ij}^{k}=1$),
this fixes $\bar\varphi^{ij}_k$ uniquely. Then
\be
 \label{eq:defect-F-def2}
\left\langle
\begin{tikzpicture}[scale=1.5,baseline]
\clip (0,0) ellipse (1.6cm and 1.2cm);
\draw[step=.25cm,gray,very thin] (-1.4,-1.4) grid (1.4,1.4);
\nicecolourscheme (0,0) ellipse (1.6cm and 1.2cm);
\draw (-0.84,0.06) node {{\scriptsize $\bar\varphi^{ij}_q$}};
\draw (0.22,-0.58) node {{\scriptsize $\bar\varphi^{qk}_l$}};
\draw (-1.09,0.36) node {{\scriptsize $D_i$}};
\draw (-0.33,0.69) node {{\scriptsize $D_j$}};
\draw (0.43,0.21) node {{\scriptsize $D_k$}};
\draw (-0.52,-0.25) node {{\scriptsize $D_q$}};
\draw (-0.19,-0.93) node {{\scriptsize $D_l$}};

\draw[very thick, color=black] (0,-0.5) -- (0,-0.9);
\draw[<-,very thick, color=black] (0,-0.9) -- (0,-2);

\draw[very thick, color=black] (-0.625,0.125) -- (-0.3125,-0.1875);
\draw[<-,very thick, color=black] (-0.375,-0.125) -- (0,-0.5);

\draw[very thick, color=black] (-1.5,1) -- (-0.9375,0.4375);
\draw[<-,very thick, color=black] (-0.9375,0.4375) -- (-0.375,-0.125);

\draw[<-,very thick, color=black] (0.625,0.125) -- (0,-0.5);
\draw[very thick, color=black] (1.5,1) -- (0.625,0.125);

\draw[<-,very thick, color=black] (-0.125,0.625) -- (-0.625,0.125);
\draw[very thick, color=black] (0.5,1.25) -- (-0.125,0.625);

\filldraw [black] (0,-0.5) circle (1.5pt);
\filldraw [black] (-0.625,0.125) circle (1.5pt);
\end{tikzpicture}
\right\rangle
=
\sum_{p \in \mathcal{I}} F^{(ijk)l}_{pq} \cdot 
\left\langle
\begin{tikzpicture}[scale=1.5,baseline]
\clip (0,0) ellipse (1.6cm and 1.2cm);
\draw[step=.25cm,gray,very thin] (-1.4,-1.4) grid (1.4,1.4);
\nicecolourscheme (0,0) ellipse (1.6cm and 1.2cm);
\draw (0.84,0.05) node {{\scriptsize $\bar\varphi^{jk}_p$}};
\draw (0.22,-0.58) node {{\scriptsize $\bar\varphi^{ip}_l$}};
\draw (-0.79,0.02) node {{\scriptsize $D_i$}};
\draw (-0.05,0.54) node {{\scriptsize $D_j$}};
\draw (0.73,0.52) node {{\scriptsize $D_k$}};
\draw (0.15,-0.06) node {{\scriptsize $D_p$}};
\draw (-0.19,-0.92) node {{\scriptsize $D_l$}};

\draw[very thick, color=black] (0,-0.5) -- (0,-0.9);
\draw[<-,very thick, color=black] (0,-0.9) -- (0,-2);

\draw[very thick, color=black] (0.625,0.125) -- (0.3125,-0.1875);
\draw[<-,very thick, color=black] (0.375,-0.125) -- (0,-0.5);

\draw[very thick, color=black] (1.5,1) -- (0.9375,0.4375);
\draw[<-,very thick, color=black] (0.9375,0.4375) -- (0.375,-0.125);

\draw[<-,very thick, color=black] (-0.625,0.125) -- (0,-0.5);
\draw[very thick, color=black] (-1.5,1) -- (-0.625,0.125);

\draw[<-,very thick, color=black] (0.125,0.625) -- (0.625,0.125);
\draw[very thick, color=black] (-0.5,1.25) -- (0.125,0.625);

\filldraw [black] (0,-0.5) circle (1.5pt);
\filldraw [black] (0.625,0.125) circle (1.5pt);
\end{tikzpicture}
\right\rangle
\ee
for the same constants $F^{(ijk)l}_{pq}$. To see this, compose \eqref{eq:defect-F-def1} from below with an appropriate dual arrangement of defect junctions, and \eqref{eq:defect-F-def2} from above, to find in both cases
\be
\left\langle
\begin{tikzpicture}[scale=1.5,baseline]
\clip (0,0) ellipse (1.6cm and 1.2cm);
\draw[step=.25cm,gray,very thin] (-1.4,-1.4) grid (1.4,1.4);
\nicecolourscheme (0,0) ellipse (1.6cm and 1.2cm);
\draw (-0.54,-0.51) node {{\scriptsize $\bar\varphi^{ij}_q$}};
\draw (0.25,-0.75) node {{\scriptsize $\bar\varphi^{qk}_l$}};
\draw (0.4,0.14) node {{\scriptsize $\varphi_{jk}^p$}};
\draw (0.02,0.53) node {{\scriptsize $\varphi_{ip}^l$}};

\draw (-0.59,0.43) node {{\scriptsize $D_i$}};
\draw (-0.19,0.05) node {{\scriptsize $D_j$}};
\draw (0.17,-0.30) node {{\scriptsize $D_k$}};
\draw (0.41,0.6) node {{\scriptsize $D_p$}};
\draw (0.-0.31,-0.71) node {{\scriptsize $D_q$}};
\draw (-0.19,-1.05) node {{\scriptsize $D_l$}};
\draw (-0.19,0.94) node {{\scriptsize $D_l$}};

\draw[very thick, color=black] (0,1.3) -- (0,1);
\draw[<-,very thick, color=black] (0,1) -- (0,0.75);

\draw[->,very thick, color=black] (0,-0.75) -- (0.375,-0.375);
\draw[very thick, color=black] (0.375,-0.375) -- (0.75,0);

\draw[->,very thick, color=black] (-0.375,-0.375) -- (0,0);
\draw[very thick, color=black] (0,0) -- (0.375,0.375);

\draw[->,very thick, color=black] (-0.75,0) -- (-0.375,0.375);
\draw[very thick, color=black] (-0.375,0.375) -- (0,0.75);

\draw[->,very thick, color=black] (0,-1.3) -- (0,-1);
\draw[very thick, color=black] (0,-1) -- (0,-0.75);

\draw[->,very thick, color=black] (0,-0.75) -- (-0.1875,-0.5625);
\draw[->,very thick, color=black] (-0.1875,-0.5625) -- (-0.75+0.1875,-0.1875);
\draw[very thick, color=black] (-0.75+0.1875,-0.1875) -- (-0.75,0);

\draw[->,very thick, color=black] (0+0.75,-0.75+0.75) -- (-0.1875+0.75,-0.5625+0.75);
\draw[->,very thick, color=black] (-0.1875+0.75,-0.5625+0.75) -- (-0.75+0.1875+0.75,-0.1875+0.75);
\draw[very thick, color=black] (-0.75+0.1875+0.75,-0.1875+0.75) -- (-0.75+0.75,0+0.75);

\filldraw [black] (-0.375,-0.375) circle (1.5pt);
\filldraw [black] (0.375,0.375) circle (1.5pt);
\filldraw [black] (0,-0.75) circle (1.5pt);
\filldraw [black] (0,0.75) circle (1.5pt);
\end{tikzpicture}
\right\rangle
=
F^{(ijk)l}_{pq} \cdot 
\left\langle
\begin{tikzpicture}[scale=1.5,baseline]
\clip (0,0) ellipse (1.6cm and 1.2cm);
\draw[step=.5cm,gray,very thin] (-1.4,-1.4) grid (1.4,1.4);
\nicecolourscheme (0,0) ellipse (1.6cm and 1.2cm);
\draw (-0.25,-0.05) node {{\scriptsize $D_l$}};
\draw[->,very thick, color=black] (0,-1.5) -- (0,0);
\draw[very thick, color=black] (0,0) -- (0,1.5);
\end{tikzpicture}
\right\rangle \, .
\ee

Let us now sketch how the defect category is described in the two models we consider.

\subsubsection*{Topological defects in rational conformal field theory}

Topological defects in rational conformal field theory and their fusion have first been studied in
\cite{Petkova:2000ip} and defect junctions have been considered in \cite{tft4,Frohlich:2006ch}.
In the formalism of \cite{tft1,tft4,Frohlich:2006ch}, one starts from the category $\mathcal{C}$ of
representations of a rational vertex operator algebra $V$ (by ``rational'' we mean that it
satisfies the finiteness conditions of \cite{Huang2005}). In $\mathcal{C}$ one chooses an algebra object $A$ (a special symmetric Frobenius algebra) and this determines an oriented open/closed conformal field theory with left/right symmetry $V$. The topological defects which are compatible with the entire symmetry $V$ are described by
the category $\mathcal{D} = A\text{-mod-}A$ 
of $A$-bimodules in $\mathcal{C}$ \cite{tft1,Frohlich:2006ch}. This is automatically a semi-simple monoidal category. Topological junction fields have
left/right conformal weight $(0,0)$. For
a junction as in \eqref{eq:many-to-many-junction}, such fields are precisely labelled by elements of the corresponding $\text{Hom}$-space in the bimodule category~\cite{tft4,Frohlich:2006ch}.

The case that will be of interest for us is when $V$ is the bosonic subalgebra of the $\mathcal N=2$ super-Virasoro vertex
operator algebra with central charge $c=3-\frac{6}{d}$ for $d \in \Z_{\ge 1}$. Let us denote
its representation category by $\mathcal{C}^{\mathcal N=2}_d$. The relevant algebra
$A$ and the computation of $A\text{-mod-}A$ are given in appendix~\ref{app:cft}. Denote
the resulting category of bimodules by $\mathcal{D}_d^{\mathcal N=2}$. One finds that in fact
\be
 \mathcal{D}_d^{\mathcal N=2} \cong \mathcal{C}^{\mathcal N=2}_d
 \quad \text{ as monoidal categories.}
\ee
The simple objects in $\mathcal{D}_d^{\mathcal N=2}$, i.\,e.\ the elementary defects, are labelled by triples
$[l,m,s]$, where $l,m,s$ are integers subject to conditions \eqref{lmstrip1} and \eqref{eq:select-identify} below.
The comparison to the Landau-Ginzburg description will only
be carried out in the subcategory of $\mathcal{D}_d^{\mathcal N=2}$ which is generated by objects labelled $[l,m,0]$ (or, equivalently, $[d-2-l,m+d,2]$, see~\eqref{eq:select-identify})
via direct sums. This is a monoidal subcategory, and we denote it by $\mathcal{D}_{d,s=0}^{\mathcal N=2}$.

\subsubsection*{Topological defects in Landau-Ginzburg models}

As mentioned above, $\mathcal N=2$ supersymmetric Landau-Ginzburg models are believed to have $\mathcal N=2$ superconformal field theories as their infrared fixed points under renormalisation group flow. Since the topological twist of an $\mathcal N=2$ superconformal field theory is equivalent to its subsector built on (anti-)chiral primaries (see e.\,g.~\cite{zhoulecturenotes}), topologically twisted Landau-Ginzburg models should be equivalent
to the chiral ring of the associated conformal field theory. 

Conformal boundary conditions of a topologically B-twisted Landau-Ginzburg model with superpotential $W$ are the objects of the category $\text{MF}^{\text{R}}(W)$ of graded matrix factorisations of $W$, and its morphisms describe boundary fields. It was argued in~\cite{br0707.0922} that the same reasoning that leads to this description of the boundary sector equally well applies to defects: between theories with superpotentials $W$ and $W'$, they are given by suitable matrix factorisations of $W-W'$, where the minus sign is due to the different relative orientation of the defect line as seen from the two theories. Furthermore, the fusion of defects is described by the tensor product of~\cite{yoshinoTP,add0401,Khovanov:2004,br0707.0922}. 

Another way to understand the minus sign is the folding trick: as matrix factorisations of $W$ are the same as modules over the curved differential graded algebra $(R,0,W)$~\cite{kl0305}, defects between theories with potentials $W$ and $W'$ should be described by $(R,0,W)$-$(R',0,W')$-bimodules. The latter may also be viewed as modules over $(R,0,W)\otimes(R,0,W')^{\text{op}}$ where the opposite curved differential graded algebra is given by $(R,0,W')^{\text{op}}=(R^{\text{op}},0,-W')$, see~\cite{Khovanov:2004,kr0405232}. 

Hence we see again that defects are modelled by matrix factorisations of $W-W'$. As in the CFT case, bimodules feature prominently in the description of defects. We shall try to systematically exploit this fact, organising defects and their fields in a category $\MFR(W,W')$ (to be defined in the next section) that is entirely treated in bimodule language. 

\medskip

Defects in a single Landau-Ginzburg model with superpotential $W = x^d$ are described by $\MFR(x^d,x^d)$, or, using the equivalence between $\C[x]$-bimodules and modules over $\C[x] \otimes_\C \C[x] \cong \C[x,y]$, by matrix factorisations of the potential $x^d-y^d$.  The simplest examples are \cite{add0401,br0707.0922}
\be
P_S=
\begin{pmatrix}
0 & \prod_{i\in S}(x-\eta^i y) \\
\prod_{i\in\{0,\ldots,d-1\}\backslash S}(x-\eta^i y) & 0
\end{pmatrix}
\ee
where $S$ is a subset of $\{0,\ldots,d-1\}$ and $\eta=\E^{2\pi\I/d}$. Let us denote by $(\MFR(W,W'))_{0}$ the subcategory of $\MFR(W,W')$ whose morphism spaces are restricted to the R-charge zero subspaces (see definitions~\ref{Rchargeofmorph} and~\ref{defMFR} for details). Let $(\mathcal{P}_{d})_{0}$ be the full subcategory of $(\MFR(x^d,x^d))_{0}$ generated via direct sums by all the $P_S$ with subsets~$S$ consisting only of consecutive numbers (modulo $d$); $(\mathcal{P}_{d})_{0}$ is also closed under taking tensor products
 \cite{br0707.0922}. We can now formulate

\begin{conjecture}\label{conjec}
$\mathcal{D}^{\mathcal N=2}_{d,s=0}\cong
(\mathcal{P}_{d})_{0}
$ 
as monoidal categories.
\end{conjecture}
On the level of objects this has been established in \cite{br0707.0922}. We will test the conjecture by also comparing morphism spaces and in particular the action of the tensor product on morphisms in certain examples. 

The evidence for conjecture \ref{conjec} is gathered by computing the relevant structure on each side separately -- in this case certain fusing matrices -- 
and then comparing the results. It would be much preferable to have a more direct and conceptual argument relating the two sides. In fact, since the defects in $\mathcal{D}^{\mathcal N=2}_{d,s=0}$ are precisely those which are transparent to $G^\pm \bar G^\pm$,%
\footnote{$G^\pm \bar G^\pm$ are the four (bosonic) combinations of the holomorphic and anti-holomorphic $\mathcal N=2$ supercurrents.} we expect more generally that there is a (non-surjective) monoidal functor from the category of $G^\pm \bar G^\pm$-preserving topological defects in an $\mathcal N=2$ superconformal field theory to $(\text{MF}^{\text{R}}_{\text{bi}}(W,W))_{0}$ for an appropriate superpotential~$W$. 

\medskip

This paper is organised as follows. In section~\ref{sec:mbf} we define matrix bi-factorisations and recall the definition of a monoidal category. Then we show that matrix bi-factorisations of arbitrary polynomial superpotentials are endowed with a monoidal structure. In section~\ref{sec:apps} we restrict to the one-variable case and compare the monoidal structure of graded matrix bi-factorisations to that of the associated minimal superconformal field theory. Some details of the proofs of section~\ref{sec:mbf} and the CFT computations of section~\ref{sec:apps} are postponed to the appendix.

\section{Matrix bi-factorisations}\label{sec:mbf}

In this section we define matrix bi-factorisations of arbitrary superpotentials and show that they form a monoidal category as expected from their physical interpretation as defects in Landau-Ginzburg models. After setting up a suitable bimodule language we discuss the case of superpotentials in only one variable in detail, treating both the situation with and without R-charge. Then we generalise these results to any finite number of variables. 

\subsection{Preliminaries}\label{prelims}

Let $k$ be a field, let $R_1$ and $R_2$ be commutative $k$-algebras and let $W_1\in R_1$, $W_2\in R_2$ be two
elements that we call superpotentials. Every $R_1$-$R_2$-bimodule $M$ comes equipped with a left action $\rho^l_M:R_1\times M\rightarrow M$ and a right action $\rho^r_M:M\times R_2\rightarrow M$. We call an $R_1$-$R_2$-bimodule free if the corresponding $(R_1\otimes_k  R_2)$-left module is free.

\begin{definition}\label{defMFbi}
A \textit{matrix bi-factorisation of $W_1$ and $W_2$} is a tuple $D=(D_0,D_1,d_0,d_1)$
where $D_i$ are free $R_1$-$R_2$-bimodules and $d_0:D_0\rightarrow D_1$ and $d_1:D_1\rightarrow D_0$ are bimodule maps such that
\begin{align*}
(d_1\circ d_0)(m_0) & = \rho^l_{D_0}(W_1,m_0) - \rho^r_{D_0}(m_0,W_2) \, , \\
(d_0\circ d_1)(m_1) & = \rho^l_{D_1}(W_1,m_1) - \rho^r_{D_1}(m_1,W_2)
\end{align*}
for all $m_i\in D_i$.
\end{definition}
We will often simply write $D=(D_0,D_1,d_0,d_1)$ in matrix form $(\begin{smallmatrix}0&d_1\\ d_0&0\end{smallmatrix})$ if the source and target modules are clear from the context. 

\begin{definition}
The category $\DG(W_1,W_2)$ has matrix bi-factorisations of $W_1$ and $W_2$ as objects, and its morphism spaces are $\Z_2$-graded: even morphisms $\phi\equiv(\begin{smallmatrix}\phi_0&0\\ 0&\phi_1\end{smallmatrix})$ from $D$ to $D'$ are pairs of bimodule maps $\phi_0:D_0\rightarrow D'_0$, $\phi_1:D_1\rightarrow D'_1$, and odd morphisms $\psi\equiv(\begin{smallmatrix}0&\psi_1\\ \psi_0&0\end{smallmatrix})$ are pairs of bimodule maps $\psi_0:D_0\rightarrow D'_1$, $\psi_1:D_1\rightarrow D'_0$. The composition in $\DG(W_1,W_2)$ is given by matrix multiplication. 
\end{definition}
The category $\DG(W_1,W_2)$ is differential $\Z_2$-graded with the differential given by
\begin{align*}
\delta : \text{Hom}_{\DG(W_1,W_2)}(D,D') & \longrightarrow \text{Hom}_{\DG(W_1,W_2)}(D,D') \, , \\
\phi & \longmapsto D' \phi - (-1)^{|\phi|} \phi D
\end{align*}
for homogeneous $\phi$ whose $\Z_2$-degree $|\phi|$ is 0 or 1 depending on whether $\phi$ is even or odd as a supermatrix. In particular, one may consider the homotopy category $H_\delta^0\left(\DG(W_1,W_2)\right)$ whose objects are the same as those of $\DG(W_1,W_2)$ and whose morphism spaces are the zeroth $\delta$-cohomologies of the morphism spaces of $\DG(W_1,W_2)$. 

\begin{definition}
The \textit{category of matrix bi-factorisations of $W_1$ and $W_2$} is the homotopy category
\be
\MF(W_1,W_2)=H_\delta^0(\DG(W_1,W_2)) \, .
\ee
\end{definition}

Concretely this means that morphisms from $D$ to $D'$ in $\MF(W_1,W_2)$ are even matrices~$\phi$ satisfying $D'\phi=\phi D$ modulo matrices of the form $D'\psi+\psi D$ for any odd matrix~$\psi$.

Given an $R_1$-$R_2$-bimodule $M$ we can construct a new bimodule ${}_{\sigma_1}\!M_{\sigma_2}$ as follows. Let $\sigma_i\in\text{Aut}(R_i)$ be automorphisms of $R_i$. Then as sets we have ${}_{\sigma_1}\!M_{\sigma_2}=M$, but the left and right actions are twisted via $\sigma_1$ and $\sigma_2$, respectively: 
\be
\rho^l_{{}_{\sigma_1}\!M_{\sigma_2}}(r_1,m) = \rho^l_M(\sigma_1(r_1),m) \, , \quad \rho^r_{{}_{\sigma_1}\!M_{\sigma_2}}(m,r_2) = \rho^r_M(m,\sigma_2(r_2))
\ee
for all $r_i\in R_i$ and $m\in M$. We note that one can canonically identify
\be\label{sternstern}
\text{Hom}_{R_1\text{-mod-}R_2}(M,M') = \text{Hom}_{R_1\text{-mod-}R_2}({}_{\sigma_1}\!M_{\sigma_2},{}_{\sigma_1}\!{M'}_{\sigma_2})
\ee
as the underlying maps of sets are the same. Hence for every matrix bi-factorisation~$D$ of $W_1$ and $W_2$ and automorphisms $\sigma_i$ as above, it follows that
\be
{}_{\sigma_1}\!D_{\sigma_2} = \left({}_{\sigma_1}\!(D_0)_{\sigma_2}, {}_{\sigma_1}\!(D_1)_{\sigma_2}, d_0, d_1\right)
\ee
is a matrix bi-factorisation~of $\sigma_1(W_1)$ and $\sigma_2(W_2)$. In particular, if the automorphisms are symmetries of the superpotentials, i.\,e.~$\sigma_1(W_1)=W_1$ and $\sigma_2(W_2)=W_2$, then ${}_{\sigma_1}\!D_{\sigma_2}\in\MF(W_1,W_2)$.

\subsection{Tensor structure for one variable}\label{tensorstructure}

We wish matrix bi-factorisations to describe defects in a given topological Landau-Ginzburg model with superpotential $W\in R:=\C[x]$. Thus we restrict to the case $R_1=R_2=R$ with $W_1=W_2=W$ and consider matrix bi-factorisations of $W$ and $W$, or simply of $W$ for short. We also write $\MF(W):=\MF(W,W)$. 

Under the identification of $R$-bimodules with $(R\tc  R)$-left modules, a bimodule describing a defect condition corresponds to a left module describing a boundary condition in the folded model. Because of the latter's geometric origin as superbundles~\cite{kl0210, l0312} we also only consider free bimodules in the case of defects.

The main goal of this section is to prove that $\MF(W)$ is a monoidal category, hence we now recall the definition of the latter.

\begin{definition}\label{def:monoidal}
A \textit{monoidal category} is a tuple $(\mathcal M, \otimes, \alpha, I, \lambda, \rho)$ such that
\begin{itemize}
\item $\mathcal M$ is a category, 
\item $\otimes : \mathcal M\times\mathcal M\rightarrow\mathcal M$ is a bifunctor called the \emph{tensor product},
\item $\alpha_{A,B,C}: (A\otimes B)\otimes C\rightarrow A\otimes(B\otimes C)$ are isomorphisms called \emph{associators}, natural in $A,B,C$,
\item $I\in\mathcal M$ is called the \textit{unit object},
\item $\lambda_A: I\otimes A\rightarrow A$, $\rho_A: A\otimes I\rightarrow A$ are isomorphisms, natural in $A$, called {\em left} and {\em right unit isomorphisms},
\item the \textit{triangle axiom} holds, i.\,e.~the diagramme
 \be
\xymatrix@!0{(A\otimes I)\otimes B \ar[rrrrrr]^{\alpha_{A,I,B}} \ar[rrrdd]_{\rho_A\otimes\text{id}_{B}} &&&&&& A\otimes(I\otimes B) \ar[ddlll]^{\text{id}_A\otimes\lambda_{B}} \\
&&&&&&\\
&&& A\otimes B &&&%
}
\ee
commutes for all $A,B\in\mathcal M$,
\item the \textit{pentagon axiom} holds, i.\,e.~the diagramme
\be
\xymatrix@C=0.5cm{
&(A\otimes (B\otimes C))\otimes D\ar[rd]^{\alpha_{A,B\otimes C,D}}& \\
((A\otimes B)\otimes C)\otimes D \ar[ru]^{\alpha_{A,B,C}\otimes \text{id}_{D}\;\;\;\;\;\;\;\;} \ar[d]^{\alpha_{A\otimes B,C,D}}&&A\otimes ((B\otimes C)\otimes D) \ar[d]_{\text{id}_{A}\otimes\alpha_{B,C,D}} \\
(A\otimes B)\otimes (C\otimes D)\ar[rr]^{\alpha_{A,B,C\otimes D}}&&A\otimes (B\otimes (C\otimes D))
}
\ee
commutes for all $A,B,C,D\in\mathcal M$.
\end{itemize}
\end{definition}

The tensor product $\otimes$ of $\MF(W)$ is given by~\cite{yoshinoTP,add0401,Khovanov:2004}
\begin{align}
D\otimes D' = & \left(\vphantom{{\textstyle \begin{pmatrix}d_0\otimes_R\one_{D'_0} & -\one_{D_1}\otimes_R d'_1\\\one_{D_0}\otimes_R d'_0 & d	_1\otimes_R \one_{D'_1}\end{pmatrix}}}D_0\otimes_R D'_0 \oplus D_1\otimes_R D'_1, D_1\otimes_R D'_0 \oplus D_0\otimes_R D'_1,\right. \nonumber \\
&\left. {\textstyle \begin{pmatrix}d_0\otimes_R\one_{D'_0} & -\one_{D_1}\otimes_R d'_1\\\one_{D_0}\otimes_R d'_0 & d_1\otimes_R \one_{D'_1}\end{pmatrix}}, {\textstyle \begin{pmatrix}d_1\otimes_R\one_{D'_0} & \one_{D_0}\otimes_R d'_1\\-\one_{D_1}\otimes_R d'_0 & d_0\otimes_R \one_{D'_1}\end{pmatrix}} \right) , \label{stern} \\
\phi\otimes \phi'= & \begin{pmatrix}
\phi_0\otimes_R \phi'_0 &0&0&0 \\
0&\phi_1\otimes_R \phi'_1&0&0 \\
0&0&\phi_1\otimes_R \phi'_0&0 \\
0&0&0&\phi_0\otimes_R \phi'_1
\end{pmatrix} , \label{morphtensor}
\end{align}
and it is also well-defined in $\DG^0(W)$. If we view matrix bi-factorisations as matrices then this definition of the tensor product is the natural one as it lets matrix bi-factorisations act as a derivation: 
\be
(D\otimes D')(m\tr m') = D(m)\tr m' + (-1)^{|m|} m\tr D'(m')
\ee
for all homogeneous $m\in D_0\oplus D_1$ and $m'\in D_0'\oplus D_1'$.

For three matrix bi-factorisations $D,D',D''$ of $W$ we have an isomorphism
\be
\alpha_{D,D',D''}: (D\otimes D')\otimes D''\longrightarrow D\otimes(D'\otimes D'')
\ee
in $\text{MF}_{\text{bi}}(W)$ with
\begin{subequations}\label{alpha}
\begin{align}
(\alpha_{D,D',D''})_0 & = \begin{pmatrix}
\one_{D_0\tr D'_0\tr D''_0} &0&0&0\\
0&0&0&\one_{D_0\tr D'_1\tr D''_1}\\
0&\one_{D_1\tr D'_1\tr D''_0}&0&0\\
0&0&\one_{D_1\tr D'_0\tr D''_1}&0
\end{pmatrix}, \\
(\alpha_{D,D',D''})_1 & = \begin{pmatrix}
\one_{D_1\tr D'_0\tr D''_0} &0&0&0\\
0&0&0&\one_{D_1\tr D'_1\tr D''_1}\\
0&\one_{D_0\tr D'_1\tr D''_0}&0&0\\
0&0&\one_{D_0\tr D'_0\tr D''_1}&0
\end{pmatrix}.
\end{align}
\end{subequations}

\begin{lemma}\label{lem:assopenta}
The maps $\alpha_{D,D',D''}$, viewed as morphisms in $\text{DG}_{\text{bi}}(W)$, satisfy the pentagon axiom as the associator for $\otimes$.
\end{lemma}
\begin{proof}
Matrix multiplication.
\end{proof}

We also need to identify a unit object in $\text{MF}_{\text{bi}}(W)$. To do so, we first introduce some notation to calculate with free $R$-bimodules. Every free $R$-bimodule is isomorphic to $R\tc  U\tc  R$ for some $\C$-vector space $U$. For two vector spaces $U,V$ consider a linear map $\phi \in\text{Hom}(U,V[a,b])$, where $V[a,b]$ are polynomials  in two formal variables $a,b$ with coefficients in $V$. We can write $\phi=\sum_{m,n}\phi_{mn}a^mb^n$, where $\phi_{mn}\in\text{Hom}(U,V)$. The sum may be infinite, but for a given $u \in U$ only a finite number of $\phi_{mn}(u)$ will be non-zero. From $\phi$ we obtain an $R$-bimodule map by setting
\begin{align}
\hat\phi: R\tc  U\tc  R & \longrightarrow R\tc  V\tc  R \, , \nonumber \\
r\tc  u \tc  s & \longmapsto \sum_{m,n}rx^m \tc  \phi_{mn}(u)\tc  x^ns \, . \label{bimap}
\end{align}
This gives an isomorphism 
\begin{align}
\text{Hom}(U,V[a,b]) \stackrel{\sim}{\longrightarrow} \text{Hom}_{R\text{-mod-}R}(R\tc  U\tc  R ,  R\tc  V\tc  R)\, .
\end{align}
We denote its inverse by $(\check~)$, i.\,e.~for a given $R$-bimodule map $\xi:R\tc  U\tc  R \rightarrow R\tc  V\tc  R$ we have $\xi=[\check\xi(a,b)]\hat~$. 
For long expressions we write $\widehat{\text{long}}=[\text{long}]\hat~$. For example, $[a-b]\hat~ : R \tc R \rightarrow R \tc R$ (here $U=V=\C$) is the map $r \tc s \mapsto xr \tc s - r \tc xs$, and $[a^i x^j b^k]\hat~$ from $R \tc R \tc R$ to itself (here $U=V=R$) is the map $r \tc s \tc t \mapsto x^i r \tc x^j s \tc x^k t$.
We define the composition of $\phi \in \text{Hom}(U,V[a,b])$ and $\psi \in \text{Hom}(T,U[a,b])$ to be
\begin{align} \label{eq:compose-ab-polys}
  \phi \circ \psi = \sum_{k,l,m,n} \phi_{kl} \circ \psi_{mn} \, a^{k+m} b^{l+n}
  \in \text{Hom}(T,V[a,b]) \, .
\end{align}
With this definition one has $\hat\phi \circ \hat\psi = [\phi \circ \psi]\hat~$.
For example, for $T=U=V=\C$, $\phi(a,b)$ and $\psi(a,b)$ are just polynomials, and composition amounts to multiplying these polynomials, $\hat\phi \circ \hat\psi = [\phi(a,b) \cdot \psi(a,b)]\hat~$.

We are now in a position to define the matrix bi-factorisation 
\begin{subequations}\label{unit}
\be
I = \left( R\tc R, R\tc R, \iota_0, \iota_1\right)
\ee
where
\be
\quad \iota_0 = \left[(W(a)-W(b))/(a-b)\right]\!\hat~ \, , \quad \iota_1 = [a-b]\hat~
\ee
\end{subequations}
as in \cite{Khovanov:2004,kr0405232,br0707.0922}. We will see that $I$ is the unit object in $\MF(W)$. 

Let $\mu:R\tc R\rightarrow R$ be the multiplication map, i.\,e.~$\mu(r\tc s)=rs$ for all $r,s\in R$. Then it is easy to check that
\be\label{mu}
\mu\circ \iota_0 = \partial W \mu \, , \quad \mu\circ \iota_1 = 0 \, .
\ee
For example, $(\mu\circ[a-b]\hat~)(r\tc  s)=\mu(rx\tc  s-r\tc  xs)=rxs-rxs=0$, and similarly for the other identity. 

For any $D\in\text{MF}_{\text{bi}}(W)$ we have maps
\begin{subequations}\label{lambdarho}
\begin{align}
\lambda_D = \begin{pmatrix}
\mu\otimes_R\one_{D_0}&0&0&0\\
0&0&0&\mu\otimes_R\one_{D_1}
\end{pmatrix} : I\otimes D \longrightarrow D \, , \\
\rho_D = \begin{pmatrix}
\one_{D_0}\otimes_R\mu&0&0&0\\
0&0&\one_{D_1}\otimes_R\mu&0
\end{pmatrix} : D\otimes I \longrightarrow D \, .
\end{align}
\end{subequations}
Since we have $I_0=I_1=R\tc \C \tc  R\equiv R\tc  R$, the multiplication $\mu$ can act on $I_0,I_1$. In a different guise and in another context such maps have already appeared in~\cite[props.\,15\,\&\,17]{Khovanov:2004}. 

To see that $\lambda_D,\rho_D$ are indeed morphisms in $\text{MF}_{\text{bi}}(W)$, we have to check that
\be\label{conds}
D\lambda_D=\lambda_D(I\otimes D) \, , \quad D\rho_D=\rho_D(D\otimes I) \, .
\ee
In the case of the left unit map $\lambda_D$ we have according to~\eqref{stern} and~\eqref{unit}: 
\be
I\otimes D = \begin{pmatrix}
0&0&\iota_1\otimes_R\one_{D_0}&\one_{I_0}\otimes_R d_1 \\
0&0&-\one_{I_1}\otimes_R d_0&\iota_0\otimes_R\one_{D_1} \\
\iota_0\otimes_R\one_{D_0}&-\one_{I_1}\otimes_R d_1&0&0 \\
\one_{I_0}\otimes_R d_0&\iota_1\otimes_R\one_{D_1}&0&0
\end{pmatrix}
\ee
and thus
\begin{align*}
D\lambda_D & = \begin{pmatrix}
0&0&0&d_1\circ(\mu\otimes_R\one_{D_1}) \\
d_0\circ(\mu\otimes_R\one_{D_0})&0&0&0
\end{pmatrix} , \\
(\lambda_D(I\otimes D))_0 & =
\Big( (\mu\otimes_R\one_{D_1})\circ(\one_{I_0}\otimes_R d_0)\;\;\;(\mu\otimes_R\one_{D_1})\circ(\iota_1\otimes_R\one_{D_1}) \Big)
\\
(\lambda_D(I\otimes D))_1 & =
\Big( (\mu\otimes_R\one_{D_0})\circ(\iota_1\otimes_R\one_{D_0})\;\;\;(\mu\otimes_R\one_{D_0})\circ(\one_{I_0}\otimes_R d_1) \Big) . 
\end{align*}
Using the second relation of~\eqref{mu} on these expressions one immediately verifies that $D\lambda_D=\lambda_D(I\otimes D)$ and hence $\lambda_D$ is a morphism. Similarly, one also finds that $\rho_D$ is a morphism. 

We will now show that $(I,\lambda,\rho)$ provide $(\MF(W),\otimes,\alpha)$ with the remaining structure of a monoidal category. This means that we need to prove that $\lambda_D$ and $\rho_D$ are isomorphisms, that $\alpha, \lambda, \rho$ are natural isomorphisms, and that they satisfy the triangle axiom. The latter is straightforward: 

\begin{lemma}\label{lem:triangle}
For all $D,D'\in\text{MF}_{\text{bi}}(W)$, the diagramme
\be
\xymatrix@!0{(D\otimes I)\otimes D' \ar[rrrrrr]^{\alpha_{D,I,D'}} \ar[rrrdd]_{\rho_D\otimes\text{id}_{D'}} &&&&&& D\otimes(I\otimes D') \ar[ddlll]^{\text{id}_D\otimes\lambda_{D'}} \\
&&&&&&\\
&&& D\otimes D' &&&
}
\ee
commutes (already on the DG level, i.\,e.~in~$\text{DG}_{\text{bi}}(W)$).
\end{lemma}
\begin{proof}
This follows from the definitions \eqref{morphtensor},~\eqref{alpha},~\eqref{lambdarho} and explicit matrix multiplication. 
\end{proof}

However, $\lambda$ and $\rho$ do not give isomorphisms in $\DG(W)$ and we have to pass to the cohomology level. 

\begin{lemma}\label{lem:isos}
In $\MF(W)$, $\lambda_D$ and $\rho_D$ are isomorphisms for all matrix bi-factorisations $D$, and we have $\lambda_I=\rho_I$.
\end{lemma}
The construction of inverses to $\lambda$ and $\rho$ in $\MF(W)$ is rather tedious and we relegate the proof of the above lemma to appendix~\ref{appA}, where we also specify the associated homotopies. 

\begin{theorem}\label{thm:MF}
The category $\text{MF}_{\text{bi}}(W)$ together with the structure $(\otimes,\alpha,I,\lambda,\rho)$ is monoidal.
\end{theorem}
\begin{proof}
The naturality of $\alpha$, $\lambda$ and $\rho$ and the functoriality of $\otimes$ already hold in $\DG(W)$ as is easily checked from the definitions via direct computation. Then the theorem follows from lemmas~\ref{lem:assopenta}, \ref{lem:triangle} and \ref{lem:isos}.
\end{proof}
We remark that one may also consider the full subcategory $\MF(W)_{\text{f}}$ of $\MF(W)$ consisting of all matrix bi-factorisations $D \in \MF(W)$ which are isomorphic (in $\MF(W)$) to a matrix bi-factorisation $D'$ for which the $R$-bimodules $D_0'$ and $D_1'$ have finite rank.
As shown in~\cite[prop.\,13]{Khovanov:2004} and \cite[sec.\,4.2]{br0707.0922} this subcategory is closed under taking tensor products, and hence the above theorem also holds for $\MF(W)_{\text{f}}$.

\begin{lemma}[\cite{Khovanov:2004,kr0405232,br0707.0922}]\label{IdSpec}
$\text{Hom}_{\text{MF}_{\text{bi}}(W)}(I,I)\cong R/(\partial W)$.
\end{lemma}
\begin{proof}
Let $\phi$ be a representative of an element in $\text{Hom}_{\text{MF}_{\text{bi}}(W)}(I,I)$. We can write $\phi=(\begin{smallmatrix}\hat \phi_0&0\\0&\hat \phi_1\end{smallmatrix})$ where according to the relation~\eqref{bimap}, $\phi_0,\phi_1 \in \text{Hom}(\C,\C[a,b])$ are polynomials in the formal variables $a,b$. From the closedness condition of $\phi$, i.\,e.~$I\phi=\phi I$, we find that $(a-b)\phi_1=\phi_0(a-b)$ and hence we must have $\phi_0=\phi_1$. The condition $\iota_1 \circ \hat\phi_1 = \hat\phi_0 \circ \iota_1$ is then automatic, so that the space of closed elements in $\text{Hom}_{\text{DG}^0_{\text{bi}}(W)}(I,I)$ is isomorphic to $\C[a,b]$.

Exact elements are of the form $I\psi+\psi I$ for arbitrary degree-one morphisms $\psi=(\begin{smallmatrix}0&\hat \psi_1\\ \hat \psi_0&0\end{smallmatrix})\in\text{Hom}_{\text{DG}^1_{\text{bi}}(W)}(I,I)$, and this is equal to
\be
\begin{pmatrix}
\hat \phi'_0&0\\
0&\hat \phi'_0
\end{pmatrix} , \quad \phi'_0= (a-b)\psi_0 + \psi_1\,\frac{W(a)-W(b)}{a-b} \, .
\ee
So we see that the space of exact elements in $\text{Hom}_{\text{DG}^0_{\text{bi}}(W)}(I,I)$ is isomorphic to the ideal $(a-b,(W(a)-W(b))/(a-b))$, and hence $\text{Hom}_{\text{MF}_{\text{bi}}(W)}(I,I)\cong R/(\partial W)$.
\end{proof}

\begin{remark}
The physical interpretation of this is that the space of defect fields on the identity defect is the same as the space of bulk fields in the full CFT, and that this in particular also holds for the ring of chiral primaries. The latter is equivalent to the closed sector of the associated topologically twisted Landau-Ginzburg model and hence given by the Jacobi ring $R/(\partial W)$~\cite{v1991}. 
\end{remark}

We close this subsection with another look at automorphisms which allows us to introduce the matrix bi-factorisations that will be of interest in section~\ref{sec:apps}. Any automorphism of $R=\C[x]$ is fixed by its action on $x$, so we have a family $\{\sigma_\alpha\}_{\alpha\in\C}\subset\text{Aut}(R)$ given by
\be
\sigma_\alpha(x) = \E^{\I q_x\alpha} x
\ee
where the constant $q_x$ is conventional. The infrared fixed point of a Landau-Ginzburg model with superpotential $W=x^d+\sum_{i>d}c_ix^i$ is the same as that of a model with superpotential $x^d$. Accordingly, the categories $\MF(W)$ and $\MF(x^d)$ are equivalent, and we may restrict to superpotentials of the form $W=x^d$. In this case we set 
\be
q_x=\frac{2}{d}
\ee  
such that $\sigma_\alpha(W)=\E^{2\I\alpha}W$. 

For a subset $S\subset\{0,\ldots,d-1\}$ we define
\be
p_S(a,b) = \prod_{i\in S}(a-\eta^i b)
\, , \quad  \eta = \E^{2 \pi \I /d} \, .
\ee
Then we get a matrix bi-factorisation 
\be
P_S = \left(R\tc R, R\tc R, \hat p_{\{0,\ldots,d-1\}\backslash S}, \hat p_S \right) \equiv
\begin{pmatrix}
0 & \hat p_S \\
\hat p_{\{0,\ldots,d-1\}\backslash S} & 0 
\end{pmatrix}
\ee
as in~\cite{add0401,br0707.0922}.

In the above formulation the unit object $I$ is given by $P_{\{0\}}$. Furthermore, we note that $\sigma_{\pi m}(W)=W$ and $\sigma_{\pi m}=\sigma_{\pi(m+d)}$ for all $m\in\Z$. 

\begin{lemma}
${}_{\sigma_{\pi m}}\!I_{\sigma_{\pi n}} \cong P_{\{m-n\}}$ in $\DG(x^d)$.
\end{lemma}
\begin{proof}
Recalling the notation introduced at the end of subsection~\ref{prelims}, this can be seen by direct computation: An isomorphism $\phi \equiv (\begin{smallmatrix}\phi_0&0\\0&\phi_{1}
\end{smallmatrix}): {}_{\sigma_{\pi m}}\!I_{\sigma_{\pi n}} \rightarrow P_{\{m-n\}}$ is provided by $\phi_0(r \tc s) = \sigma_{-\pi m}(r) \tc \sigma_{-\pi n}(s)$ and $\phi_1(r \tc s) = \eta^{-m} \sigma_{-\pi m}(r) \tc \sigma_{-\pi n}(s)$.
\end{proof}

\subsection{R-charge}\label{rcharge}

To compare the defect category $\MF(W)$ with the CFT description it is crucial to have additional information on the R-charges of morphisms. We will now formulate the notion of R-charge in bimodule language and indicate its relation to the standard formulation. In the following we abbreviate ${}_\alpha\!M_\alpha\equiv{}_{\sigma_\alpha}\!M_{\sigma_\alpha}$. 

\begin{definition}
A \textit{bimodule with $\mathfrak{u}(1)$-action} is a pair $(M,\varphi)$ where $M$ is an $R$-bimodule and $\varphi$ assigns to each $\alpha\in\C$ a bimodule isomorphism $\varphi_\alpha:M\rightarrow{}_\alpha\!M_\alpha$ such that $\varphi_{\alpha+\beta}=\varphi_\beta\circ\varphi_\alpha$. 
\end{definition}

Here, $\varphi_\beta$ above is understood as an element of $\text{Hom}({}_\alpha\!M_\alpha,{}_{\alpha+\beta}\!M_{\alpha+\beta})=\text{Hom}(M,{}_\beta\!M_\beta)$ via the identification~\eqref{sternstern}. Similarly, in the next definition the map $F$ is viewed as an element of both $\text{Hom}(M,M')$ and $\text{Hom}({}_\alpha\!M_\alpha,{}_\alpha\!{M'}_\alpha)$.

\begin{definition}\label{Rchargeofmorph}
Let $(M,\varphi)$ and $(M',\varphi')$ be bimodules with $\mathfrak{u}(1)$-action and let $F:M\rightarrow M'$ be a bimodule map. We define
\be\label{Falpha}
F_\alpha = (\varphi'_\alpha)^{-1}\circ F\circ\varphi_\alpha \, ,
\ee
and we say that $F$ \textit{has R-charge $q$} if $F_\alpha=\E^{-\I\alpha q}F$ for all $\alpha\in\C$. 
\end{definition}

The minus sign in $\E^{-\I\alpha q}$ is used to match the standard conventions in the literature as we will see below. 

\begin{definition}
A \textit{graded matrix bi-factorisation} is a tuple
\be
D=(D_0,D_1,d_0,d_1,\varphi^{D_0},\varphi^{D_1})
\ee	
such that
\begin{itemize}
 \item $(D_0,D_1,d_0,d_1)$ is a matrix bi-factorisation,
 \item $(D_0,\varphi^{D_0})$ and $(D_1,\varphi^{D_1})$ are bimodules with $\mathfrak u(1)$-action,
 \item $d_0$ and $d_1$ have R-charge 1. 
\end{itemize}
\end{definition}
To make contact with the standard formulation, we first define for any $\C$-vector space $V$ the bimodule maps
\begin{align}
s_\alpha^V: R\tc V\tc R & \longrightarrow {}_\alpha\! (R\tc V\tc R)_\alpha \, , \nonumber \\
r\tc v\tc s & \longmapsto \sigma_\alpha(r)\tc v \tc\sigma_\alpha(s) \, .
\end{align}
Then for a matrix bi-factorisation $D$ with $D_0 = R \tc \check D_0 \tc R$ and $D_1 = R \tc \check D_1 \tc R$ for some vector spaces $\check D_0$, $\check D_1$, we set
\be
U_{D_i}(\alpha) = \left(\varphi_{-\alpha}^{D_i}\right)^{-1}\circ s_{-\alpha}^{\check D_i} \, , \quad U_D(\alpha)=\begin{pmatrix}
U_{D_0}(\alpha) & 0 \\
0 & U_{D_1}(\alpha)
\end{pmatrix} .
\ee
Now let $\hat\phi:D\rightarrow D'$ be a morphism of graded matrix bi-factorisations. It follows from the above and equation~\eqref{Falpha} that the condition of $\hat\phi$ having R-charge $q$ is equivalent to
\be
U_{D'}(\alpha)\, [\phi(\sigma_\alpha(a), \sigma_\alpha(b))]\hat~ \left(U_D(\alpha)\right)^{-1} = \E^{+\I q \alpha} [\phi(a,b)]\hat~
\ee
for all $\alpha\in\C$, which is the standard R-charge condition.

\begin{definition}\label{defMFR}
The \textit{category of graded matrix bi-factorisations} $\MFR(W)$ has graded matrix bi-factorisations as objects and the same morphisms as $\MF(W)$.
\end{definition}

The tensor product of $\MFR(W)$ is the same as that of $\MF(W)$ apart from the fact that we have to specify what $\otimes$ does to the $\mathfrak{u}(1)$-action: for the tensor product $(D_0,D_1,d_0,d_1,\varphi^{D_0},\varphi^{D_1})\otimes(D'_0,D'_1,d'_0,d'_1,\varphi^{D'_0},\varphi^{D'_1})$ of two graded matrix bi-factorisations it is given by
\be
((\varphi\otimes\varphi')_0, (\varphi\otimes\varphi')_1) \quad\text{with}\quad \varphi=
\begin{pmatrix}
\varphi^{D_0} & 0 \\
0 & \varphi^{D_1}
\end{pmatrix}, \quad
\varphi'=
\begin{pmatrix}
\varphi^{D'_0} & 0 \\
0 & \varphi^{D'_1}
\end{pmatrix}
\ee
where we use~\eqref{morphtensor} to compute $\varphi\otimes\varphi'$. 

We also have to specify the $\mathfrak u(1)$-action of the unit object. It is given by
\be
U_I(\alpha) =
\begin{pmatrix}
1 & 0 \\
0 & \E^{\I\alpha(q_x-1)}
\end{pmatrix} .
\ee
We will see in section~\ref{sec:apps} that this is a special case of the general R-charge matrices $U_{P_S}$ which we will infer from comparison with the conformal field theory description. 

It is easy to compute the charges of the spectrum of the identity defect:
\begin{lemma}
$\text{Hom}_{\MFR(x^d)}(I,I)=\C\{\phi_I^i\}_{i\in\{0,\ldots,d-2\}}$, where $\phi_I^i=
(\begin{smallmatrix}
\hat a^i & 0 \\
0 & \hat a^i
\end{smallmatrix})$ has R-charge~$2i/d$. 
\end{lemma}
\begin{proof}
This follows as in lemma \ref{IdSpec} and from a direct computation. 
\end{proof}

Similarly, the associator and the left and right unit isomorphisms have zero R-charge as expected: 
\begin{lemma}
$\alpha_{D,D',D''}$, $\lambda_D$ and $\rho_D$ have R-charge 0 for all $D,D',D''$ in $\MFR(W)$.
\end{lemma}
\begin{proof}
Direct computation. 
\end{proof}

The arguments of the previous subsection immediately carry over to the graded situation and we have:
\begin{theorem}\label{thm:MFR}
$(\MFR(W),\otimes,\alpha,(I,U_I),\lambda,\rho)$ is a monoidal category.
\end{theorem}

\subsection{Tensor structure for many variables}\label{subsec:manyV}

We shall now generalise the above results to the many-variable case, i.\,e.~$W\in R=\C[x_1,\ldots,x_N]$ for arbitrary $N\in\Z_{\geq 1}$. We will use the obvious generalisation of the ``$[\,\cdot\,]\hat~$-calculus'' introduced in subsection~\ref{tensorstructure} to the case of many formal variables $a_i,b_i$. Also, in the graded case the superpotential $W$ has to be quasi-homogeneous of degree~2. 

Matrix bi-factorisations and their morphisms are defined as before, and also the definitions of the tensor product~$\otimes$ and the associator~$\alpha$ remain unchanged. The unit object is slightly more involved: as in~\cite{Khovanov:2004} we set
\begin{align}
\Delta_i(W) & = \frac{W(\boldsymbol b_{[i-1]} \boldsymbol a)-W(\boldsymbol b_{[i]} \boldsymbol a)}{a_i-b_i} \, , \quad \boldsymbol b_{[i]} \boldsymbol a = (b_1,\ldots,b_i,a_{i+1},\ldots,a_N)
\end{align}
for all $i\in \{1,\ldots,N\}$. Then the unit object is given by
\be
I = \left( (R\tc R)^{\oplus 2^{N-1}}, (R\tc R)^{\oplus 2^{N-1}}, \iota_0, \iota_1 \right)
\ee
where in matrix form we have
\be\label{eq:I-N-factor}
\begin{pmatrix}
0 & \iota_1 \\
\iota_0 & 0
\end{pmatrix}
=\sideset{}{^{{}_{'}}}\bigotimes_{i=1}^N
\begin{pmatrix}
0 & [a_i-b_i]\hat~ \\
[\Delta_i(W)]\hat~ & 0
\end{pmatrix} .
\ee
Here, the operation $\otimes'$ is defined like the tensor product in~\eqref{stern} apart from the fact that ``$\one\tr$'' and ``$\tr\one$'' do not appear in~$\otimes'$. 
For example, we have
\be
I=
\begin{pmatrix}
0 & 0 & [a_1-b_1]\hat~ & [a_2-b_2]\hat~ \\
0 & 0 &  -[\Delta_2(W)]\hat~ & [\Delta_1(W)]\hat~ \\
[\Delta_1(W)]\hat~ & -[a_2-b_2]\hat~ & 0 & 0 \\
[\Delta_2(W)]\hat~ & [a_1-b_1]\hat~ & 0 & 0
\end{pmatrix}
\quad\text{for } N=2.
\ee
For $N>2$, the $N$-fold product is to be bracketed from the left, i.\,e., denoting the $i$-th factor in \eqref{eq:I-N-factor} by $M_i$, the expression stands for $( \ldots ((M_1 \otimes' M_2) \otimes' M_3) \otimes' \ldots ) \otimes' M_N$.
The R-charge matrix $U_I$ of $I$ is given by the tensor product of the corresponding one-variable matrices. 

The left and right unit isomorphisms
\be
\lambda_D: I\otimes D \longrightarrow D \, , \quad \rho_D: D\otimes I \longrightarrow D
\ee
for a matrix bi-factorisation~$D$ are in general represented by $(2\times2^{N+1})$-block matrices. Explicitly, they are given via
\begin{subequations}\label{lambdarho-gen}
 \begin{align}
(\lambda_D)_{i,j} & = (\mu\tr\one_{D_0}) \delta_{i,1}\delta_{j,1} + (\mu\tr\one_{D_1}) \delta_{i,2}\delta_{j,3\cdot2^{N-1}+1} \, , \\
(\rho_D)_{i,j} &  = (\one_{D_0}\tr\mu) \delta_{i,1}\delta_{j,1} + (\one_{D_1}\tr\mu) \delta_{i,2}\delta_{j,2^{N}+1} \, .
\end{align}
\end{subequations}

\medskip

We recall from subsection~\ref{tensorstructure} that $\MF(W)_{\text{f}}$ is the full subcategory of $\MF(W)$ consisting of all matrix bi-factorisations which are isomorphic (in $\MF(W)$) to matrix factorisations with finite-rank $R$-bimodules.
The category $\MFR(W)_{\text{f}}$ is defined analogously, and we have the following result. 
\begin{theorem}\label{thm:MFBmany}
$(\MF(W)_{\text{f}},\otimes,\alpha,I,\lambda,\rho)$ and $(\MFR(W)_{\text{f}},\otimes,\alpha,(I,U_I),\lambda,\rho)$ are monoidal categories. 
\end{theorem}
The proof is given in appendix~\ref{appA} where we also discuss the inverses to~$\lambda_D$ and~$\rho_D$ as needed. Our proof requires the restriction to $\MF(W)_{\text{f}}$ and $\MFR(W)_{\text{f}}$, but we suspect that this is not necessary.

\section{Applications}\label{sec:apps}

In this section we examine matrix bi-factorisations of $W=x^d$ in more detail and compare them to the description of defects in the associated $\mathcal N=2$ minimal conformal field theories. We do so by computing fusing matrices and find that both descriptions produce the same results. This is expected from the Landau-Ginzburg/CFT correspondence and adds strength to conjecture~\ref{conjec}. As is familiar from other situations, the explicit matrix factorisation computations are more straightforward than the conformal field theory analysis. 

\subsection{Conformal field theory calculations}\label{subsec:CFTcalcs}

In this subsection we use the coset description of $\mathcal N=2$ minimal models to compute a number of spectra of defect fields and fusing matrices in the defect category.

\subsubsection[$\mathcal N=2$ minimal models]{$\boldsymbol{\mathcal N=2}$ minimal models}

Let $\sVir_d$ be the $\mathcal N=2$ super-Virasoro vertex operator algebra with central charge $c_d = 3-\frac{6}{d}$ for some integer $d \ge 3$, see e.\,g.~\cite{Kazama:1988qp,Eholzer:1996zi,Adamovic:1998,Huang:2000}. Its bosonic part $(\sVir_d)_\text{bos}$ can be obtained via the coset construction, the relevant coset being $\big(\widehat{\mathfrak{su}}(2)_{d-2} \oplus \widehat{\mathfrak{u}}(1)_4 \big) / \widehat{\mathfrak{u}}(1)_{2d}$. The vertex operator algebra $(\sVir_d)_\text{bos}$ contains in particular the bosonic stress tensor $T$ with zero mode $L_0$, and a $\widehat{\mathfrak{u}}(1)$-current~$J$ with zero mode $J_0$.

Denote the category of representations of $(\sVir_d)_\text{bos}$ by $\mathcal{C}_d^{\mathcal N=2}$; it is a semi-simple braided monoidal category. The simple objects in $\mathcal{C}_d^{\mathcal N=2}$, i.\,e.\ the irreducible representations of $(\sVir_d)_\text{bos}$, are labelled by triples
\be\label{lmstrip1}
    [l,m,s] \quad \text{where}\quad  l \in \{0,\dots,d-2\} \, , \; m \in \Z_{2d} \, , \; s\in\Z_{4}
\ee
subject to the selection rule and field identification
\be
 l+m+s \in 2\Z \quad \text{and} \quad [l,m,s] = [d-2-l,d+m,s+2] \, .
 \label{eq:select-identify}
\ee
Let $R_{[l,m,s]}$ be the corresponding representation.
The decomposition of the fusion tensor product of two representations
$R_{[l,m,s]}$ and $R_{[l',m',s']}$ is described by the fusion ring with
\be
 [l,m,s] \star [l',m',s'] =
 \sideset{}{^{{}_{(+2)}}}\sum_{u = |l-l'|}^{\text{min}(l+l',2d-4-l-l')} [u,m+m',s+s'] \, ,
 \label{eq:N=2-svir-rep-fusion}
\ee
where the superscript $\,(+2)\,$ means that the sum is carried out
in steps of two.
The chiral and anti-chiral primaries are those states 
in the NS sector (i.\,e.\ in representations $R_{[l,m,s]}$ with $s$ even)
for which $q = \pm 2h$,
with $q$ the $J_0$-eigenvalue of the state and $h$ its $L_0$-eigenvalue.
Chiral or anti-chiral primaries occur only in the representations
$R_{[l, \pm l,0]}$ for $l \in \{0,\dots,d-2\}$, and there
they are the highest weight states. The relevant eigenvalues are
\be\label{qcharges}
 q_{[l,l,0]} = 2 h_{[l,l,0]} = -q_{[l,-l,0]} = 2 h_{[l,-l,0]} = \frac{l}{d} \, .
\ee

We will be interested in the A-type modular invariant theory. The bosonic part of its
space of bulk fields reads
\be
 \mathcal{H}_\text{bos} = \bigoplus_{[l,m,s]}
 R_{[l,m,s]} \otimes \bar R_{[l,m,-s]} \, ,
\label{eq:N=2-mm-bulk}
\ee
where the sum runs over the set of labels for irreducible representations.
The representations containing chiral primaries are thus precisely
\be
 \mathcal{H}_\text{bos,chiral}
 = \bigoplus_{l=0}^{d-2} R_{[l,l,0]} \otimes \bar R_{[l,l,0]}
  \subset \mathcal{H}_\text{bos} \, .
\label{eq:N=2-mm-bulk-chiral}
\ee

\subsubsection[Topological defects in the A-type minimal models]{Topological defects in the A-type $\boldsymbol{\mathcal N=2}$ minimal models}

Let us denote by $\mathcal{D}_d^{\mathcal N=2}$ the monoidal category describing the topological defects in the bosonic subsector of
the A-type $\mathcal N=2$ minimal model at level $d-2$ which preserve the entire bosonic
chiral algebra $(\sVir_d)_\text{bos} \otimes (\overline{\sVir}_d)_\text{bos}$. Recall
from the introduction that
\begin{itemize}
\item the objects of $\mathcal{D}_d^{\mathcal N=2}$ label different defect conditions,
\item the tensor product describes the fusion of defect lines,
\item the morphism spaces
$\text{Hom}_{\mathcal{D}_d^{\mathcal N=2}}(X_1 \otimes \ldots \otimes X_m, Y_1 \otimes \ldots \otimes Y_n)$
describe defect-junction fields of left/right conformal weight $(0,0)$ that
sit on a junction point with $m$ incoming defect lines labelled $X_1,\dots,X_m$ and
$n$ outgoing defect lines labelled $Y_1,\dots,Y_n$.
\end{itemize}
The monoidal category $\mathcal{D}_d^{\mathcal N=2}$ can be computed using the methods of \cite{tft1,Frohlich:2003hm,Frohlich:2006ch}. This is sketched in appendix \ref{app:cft}, here we just summarise the results, namely one finds that
\be
 \mathcal{D}_d^{\mathcal N=2} \cong \mathcal{C}_d^{\mathcal N=2}
 \quad \text{as monoidal categories.}
 \label{eq:N=2-Cd=Dd}
\ee
Consequently, we can use the same set of labels $[l,m,s]$ for simple objects in $\mathcal{D}_d^{\mathcal N=2}$ as we did for $\mathcal{C}_d^{\mathcal N=2}$. To reduce confusion, let us denote the simple objects of $\mathcal{D}_d^{\mathcal N=2}$ by
$D_{[l,m,s]}$ rather than by $R_{[l,m,s]}$. Because of the equivalence \eqref{eq:N=2-Cd=Dd},
the tensor product of two such irreducible objects decomposes as
\be
D_{[l,m,s]} \otimes D_{[l',m',s']}
\cong \bigoplus_{[u,n,t]} D_{[u,n,t]} \, ,
\label{eq:Dk-tensor-product}
\ee
where the sum runs over all labels $[u,n,t]$ that occur in the fusion product $[l,m,s] \star [l',m',s']$
as given in \eqref{eq:N=2-svir-rep-fusion}.
Note that ${\bf 1} = D_{[0,0,0]}$ is the unit object in $\mathcal{D}_d^{\mathcal N=2}$; it corresponds
to the identity defect.
The fusion ring of elementary defects in $\mathcal N=2$ minimal models has also been
computed in \cite{br0707.0922}.

We can now ask about the space of 2-fold junction fields, i.\,e.~defect fields on a given defect, or defect changing fields between two distinct defects. For comparison with the matrix factorisation
computation in the next subsection, it will be useful to know the chiral primaries in the spaces of bosonic defect changing fields from the defect labelled $[0,2n,0]$ to the defect labelled $[u,u+2n,0]$. These are given in \cite{bg0503,err0508} (as boundary spectra of permutation branes), and we also compute them in the appendix, with the result that 
the space of bosonic defect changing fields $[0,2n,0] \rightarrow [u,u+2n,0]$ is
\be
 \mathcal{H}_\text{bos}^{[0,2n,0] \rightarrow [u,u+2n,0]}
 = \bigoplus_{[l,m,s],[l',m',s']} R_{[l,m,s]} \otimes \bar R_{[l',m',s']} \, ,
 \label{eq:N=2-mm-defect-change}
\ee
where the sum runs over those pairs $[l,m,s],[l',m',s']$ which have the property
that $[u,u,0]$ appears in the fusion product $[l,m,s] \star [l',-m',s']$.
In particular, the result is independent of $n$. It is easy to see that the
subspace of representations containing chiral primaries is
\be \label{eq:N=2-defch-chiral}
 \mathcal{H}_\text{bos, chiral}^{[0,2n,0] \rightarrow [u,u+2n,0]}
 = \bigoplus_{l=0}^{d-u-2} R_{[u+l,u+l,0]} \otimes \bar R_{[l,l,0]} \, .
\ee
In the special case $n=u=0$, the expression \eqref{eq:N=2-mm-defect-change}
describes the defect fields on the identity defect, which is
nothing but the space of bulk fields \eqref{eq:N=2-mm-bulk}. 

\subsubsection{Some examples of fusing matrices in the defect category}\label{subsub:cft-ex}

As described in the introduction, we can use two-to-one defect junctions
to build many-to-one defect junctions.
In the semi-simple case, we can use the two-to-one junctions to build two different bases of the
space of three-to-one junctions. These bases are related by the fusing matrices, or 6j-symbols.

For the comparison to matrix factorisations we will only use the subset of objects labelled by $[i,j,0]$.
Let $D_{[m,n,0]}$ be a simple object that appears in the decomposition
of $D_{[i,j,0]} \otimes D_{[k,l,0]}$ as given in \eqref{eq:Dk-tensor-product}. It occurs
with multiplicity one, and a basis of the corresponding morphism space is provided by choosing a non-zero
element
\be
 \lambda_{(i,j)(k,l)}^{(m,n)} \in
 \text{Hom}_{\mathcal{D}_d^{\mathcal N=2}}( D_{[i,j,0]} \otimes D_{[k,l,0]} , D_{[m,n,0]} ) \, .
 \label{eq:N=2-mm-3-Hom-basis}
\ee
Fix a non-zero $\lambda_{(i,j)(k,l)}^{(m,n)}$ in each such non-zero $\text{Hom}$-space.
We now list the fusing matrices in the category $\mathcal{D}_d^{\mathcal N=2}$ in three examples. The corresponding calculations are sketched in appendix \ref{app:cft}.

\bigskip

\nxt {\bf $\boldsymbol{(D_{[0,2i,0]} \otimes D_{[0,2j,0]}) \otimes D_{[0,2k,0]} \longrightarrow D_{[0,2i+2j+2k,0]}}$ with $\boldsymbol{i,j,k \in\Z_d}$:}

\medskip

\noindent From the fusion rules we see that this morphism space is one-dimensional. The two basis vectors obtained from combining the elements \eqref{eq:N=2-mm-3-Hom-basis} are necessarily related by a non-zero constant,
\be
 \lambda_{(0,2i)(0,2j+2k)}^{(0,2i+2j+2k)} \circ (\id\otimes \lambda_{(0,2j)(0,2k)}^{(0,2j+2k)})\circ\alpha
 = \psi_{i,j,k} \cdot
 \lambda_{(0,2i+2j)(0,2k)}^{(0,2i+2j+2k)} \circ (\lambda_{(0,2i)(0,2j)}^{(0,2i+2j)} \otimes \id) \, ,
\ee
where $\psi_{i,j,k}\in \C^\times$ and $\alpha$ denotes the associator in $\mathcal{D}_d^{\mathcal N=2}$.
The value of the constants $\psi_{i,j,k}$ depends on the choice of basis vectors
$\lambda_{(0,2i)(0,2j)}^{(0,2i+2j)}$. The pentagon identity implies that $\psi$ is a 3-cocycle on
$\Z_{d}$, and different basis choices change it by a coboundary, see e.\,g.~\cite{Moore:1988qv,tft3}. The basis independent information contained in $\psi$ is thus a class in $H^3(\Z_d,\C^\times)\cong\Z_d$. For the specific choices made in the calculation in
appendix \ref{app:cft} one simply finds
\be\label{psi1}
 \psi_{i,j,k} = 1 \qquad \text{for all } i,j,k \in \Z_d \, ,
\ee
so that  $\psi$ represents the trivial class in $H^3(\Z_d,\C^\times)$.

\bigskip

\nxt {\bf $\boldsymbol{(D_{[d-2,0,0]} \otimes D_{[f,f,0]}) \otimes D_{[d-2,0,0]} \longrightarrow D_{[f,f,0]}}$, $\boldsymbol{d\in 2\Z},\,\boldsymbol{f=\frac{d}{2}-1}$:}

\medskip

\noindent The fusion rules are $D_{[f,f,0]} \otimes D_{[d-2,0,0]} \cong D_{[f,f,0]}$, so that this morphism space is again one-dimensional, and we find
\be\label{pm1}
 \lambda_{(d-2,0)(f,f)}^{(f,f)} \circ (\id \otimes \lambda_{(f,f)(d-2,0)}^{(f,f)})\circ\alpha
 = (-1)^{(d-2)/2} \cdot
 \lambda_{(f,f)(d-2,0)}^{(f,f)} \circ (\lambda_{(d-2,0)(f,f)}^{(f,f)} \otimes \id) \, .
\ee
Because the same basis morphisms $\lambda$ appear on either side of this equation,
the factor $(-1)^{(d-2)/2}$ is independent of the choice of basis in \eqref{eq:N=2-mm-3-Hom-basis}.

\bigskip

\nxt {\bf $\boldsymbol{(D_{[1,1,0]} \otimes D_{[1,1,0]}) \otimes D_{[1,1,0]} \longrightarrow D_{[1,3,0]}}$, $\boldsymbol{d \ge 4}$:}

\medskip

\noindent This morphism space is two-dimensional, and accordingly the two bases are related by a $(2 \times 2)$-matrix $F$, namely, for $p \in \{0,2\}$,
\be\label{2x2eq}
 \lambda_{(1,1)(p,2)}^{(1,3)} \circ (\id \otimes \lambda_{(1,1)(1,1)}^{(p,2)}) \circ\alpha
 = \sum_{q \in \{0,2\}} F_{pq} \cdot
 \lambda_{(q,2)(1,1)}^{(1,3)} \circ (\lambda_{(1,1)(1,1)}^{(q,2)} \otimes \id) \, .
\ee
Under a rescaling
$\lambda_{(i,j)(k,l)}^{(m,n)} \mapsto \eta_{(i,j)(k,l)}^{(m,n)} \cdot \lambda_{(i,j)(k,l)}^{(m,n)}$ for
some $\eta_{(i,j)(k,l)}^{(m,n)} \in \C^\times$, the entries of $F$ change as
\begin{align}
 F_{00} &\longmapsto \frac{\eta_{(0,2)(1,1)}^{(1,3)}}{\eta_{(1,1)(0,2)}^{(1,3)}} F_{00} \, , &
 F_{02} &\longmapsto \frac{\eta_{(2,2)(1,1)}^{(1,3)}\eta_{(1,1)(1,1)}^{(2,2)} }{
   \eta_{(1,1)(0,2)}^{(1,3)} \eta_{(1,1)(1,1)}^{(0,2)} } F_{02} \, , \nonumber
\\
 F_{20} &\longmapsto \frac{\eta_{(0,2)(1,1)}^{(1,3)}\eta_{(1,1)(1,1)}^{(0,2)} }{
   \eta_{(1,1)(2,2)}^{(1,3)} \eta_{(1,1)(1,1)}^{(2,2)} } F_{20} \, ,&
 F_{22} &\longmapsto \frac{\eta_{(2,2)(1,1)}^{(1,3)}}{\eta_{(1,1)(2,2)}^{(1,3)}} F_{22} \, .
\end{align}
The six independent constants can be used to adjust three of the four entries of $F$ at will.
The fourth one is then fixed, because the constants $\eta_{(i,j)(k,l)}^{(m,n)}$ cancel from the ratio
\be\label{Fratio}
 \frac{F_{00} F_{22}}{F_{02} F_{20}} \, .
\ee
Provided that all entries of $F$ are non-zero,
this ratio is thus the only basis-independent information contained in $F$.
For the choice for \eqref{eq:N=2-mm-3-Hom-basis} implicit in
appendix \ref{app:cft} one has
\be\label{FCFT}
 F_{00} = - F_{22} = \frac{-1}{2\cos \frac{\pi}{d}} \, , \quad
 F_{02} = F_{20} = \frac{\sqrt{\sin \frac{\pi}{d}\sin \frac{3\pi}{d}}}{\sin \frac{2\pi}{d}}
\ee
so that
\be\label{Fratio2}
 \frac{F_{00} F_{22}}{F_{02} F_{20}} = \frac{-1}{1+2\cos \frac{2\pi}{d}} \, .
\ee

\subsection{Matrix bi-factorisation calculations}\label{subsec:MBFcalcs}

We will now reproduce the conformal field theory results of the previous subsection in the technically less pronounced matrix bi-factorisation formalism. To do so, we make use of the fact~\cite{br0707.0922} that the defect
\be\label{dict}
D_{[l,l+2m,0]} \text{ corresponds to } P_{\{m,\ldots,m+l\}}
\ee
in $\MF(x^d)$. However, in order to compare with conformal field theory the object $P_{\{m,\ldots,m+l\}}$ needs to be lifted to the graded category $\MFR(x^d)$. This is easily done by constructing R-charge matrices $U_{m,\ldots,m+l}\equiv U_{P_{\{m,\ldots,m+l\}}}$ with respect to which the morphisms in $\MFR(x^d)$ have the same charges as the defect fields in the CFT description. 

More precisely, we can check that
\be
\text{Hom}_{\MF(x^d)}(P_{\{0\}},P_{\{0,\ldots,j\}})\cong\text{Hom}_{\MF(x^d)}(P_{\{i\}},P_{\{i,\ldots,i+j\}})
\ee
for all $i,j\in\N$, and an explicit basis for $\text{Hom}_{\MF(x^d)}(P_{\{0\}},P_{\{0,\ldots,j\}})$ is given by
\be
\phi_k =
\begin{pmatrix}
[a^k p_{\{1,\ldots,j\}}(a,b)]\hat~ & 0 \\
0 & \hat a^k
\end{pmatrix} , \quad k\in\{0,\ldots,d-j-2\} \, .
\ee
But from~\eqref{qcharges} and~\eqref{eq:N=2-defch-chiral} 
we know that also the charges of the spectra $P_{\{0\}}\rightarrow P_{\{0,\ldots,j\}}$ and $P_{\{i\}}\rightarrow P_{\{i,\ldots,i+j\}}$ coincide, which means that we must have $U_{0,\ldots,j}(\alpha)=U_{i,\ldots,i+j}(\alpha)$. Then with the general ansatz $U_{0,\ldots,j}(\alpha)=
(\begin{smallmatrix}
\E^{\I q_x\alpha c_1(j)} & 0 \\
0 & \E^{\I q_x\alpha c_2(j)}
\end{smallmatrix})$ we find that the condition of $P_{\{0,\ldots,j\}}$ having R-charge~1 translates into $c_2(j)=c_1(j)+j+\frac{2-d}{2}$. The constants $c_1(j)$ can be fixed by computing the R-charges of $\phi_k$ to be $q_k=2(c_1(j)+j+k)/d$. Comparing this with~\eqref{qcharges} and~\eqref{eq:N=2-defch-chiral} 
we get $c_1(j) = -\frac{j}{2}$ and thus
\be
U_{0,\ldots,j}(\alpha)=U_{i,\ldots,i+j}(\alpha)=
\begin{pmatrix}
\E^{-\I q_x\alpha j/2} & 0 \\
0 & \E^{\I q_x\alpha(j-d+2)/2}
\end{pmatrix}.
\ee

Now we can treat the examples of the previous subsection in the Landau-Ginzburg description. 

\bigskip

\nxt {\bf $\boldsymbol{(P_{\{i\}}\otimes P_{\{j\}})\otimes P_{\{k\}} \longrightarrow P_{\{i+j+k\}}}$:}

\medskip\noindent
A possible choice of non-zero morphisms in $\MFR(x^d)$ corresponding to the defect fields $\lambda_{(0,2i)(0,2j)}^{(0,2i+2j)}$ of charge~0 between group-like defects is
\be
\Lambda_{i,j} =
\begin{pmatrix}
M_j & 0 & 0 & 0 \\
0 & 0 & M_j & 0
\end{pmatrix} :
P_{\{i\}}\otimes P_{\{j\}} \longrightarrow P_{\{i+j\}}
\ee
where  $M_j$ is the bimodule map
\be
R\tc R\tc R \longrightarrow R\tc R \, , \quad 1\tc x^m \tc 1 \longmapsto \eta^{jm}\cdot 1\tc x^m \, .
\ee
It is straightforward to check that
\be
\Lambda_{i,j+k}\circ(\id\otimes\Lambda_{j,k})\circ \alpha = 1\cdot \Lambda_{i+j,k}\circ(\Lambda_{i,j}\otimes\id)
\ee
already in $\DGR(x^d)$. Hence we recover the same fusing ``matrix''~1 as in~\eqref{psi1}.

\bigskip

\nxt {\bf $\boldsymbol{(P_{\{\frac{d}{2}+1,\ldots,\frac{3d}{2}-1\}}\otimes P_{\{0,\ldots,\frac{d}{2}-1\}})\otimes P_{\{\frac{d}{2}+1,\ldots,\frac{3d}{2}-1\}} \longrightarrow P_{\{0,\ldots,\frac{d}{2}-1\}}}$, $\boldsymbol{d\in 2\Z}$:}

\medskip\noindent
Let us next consider the twice as interesting example where the fusing ``matrix'' can be $+1$ or $-1$. According to the dictionary~\eqref{dict} there should be morphisms
\begin{align*}
F_l: \;&  P_{\{\frac{d}{2}+1,\ldots,\frac{3d}{2}-1\}}\otimes P_{\{0,\ldots,\frac{d}{2}-1\}} \longrightarrow P_{\{0,\ldots,\frac{d}{2}-1\}} \, , \\
F_r: \; & P_{\{0,\ldots,\frac{d}{2}-1\}}\otimes P_{\{\frac{d}{2}+1,\ldots,\frac{3d}{2}-1\}} \longrightarrow P_{\{0,\ldots,\frac{d}{2}-1\}}
\end{align*}
in $\MFR(x^d)$ that correspond to the defect fields $\lambda_{(f,f)(d-2,0)}^{(f,f)}$, $\lambda_{(d-2,0)(f,f)}^{(f,f)}$ in~\eqref{pm1}. A possible choice is
\be
F_l=
\begin{pmatrix}
0 & M'_{\frac{d}{2}} & 0 & 0 \\
0 & 0 & (-1)^{\frac{d}{2}-1}M'_{\frac{d}{2}} & 0
\end{pmatrix} , \quad
F_r=
\begin{pmatrix}
0 & M_{\frac{d}{2}} & 0 & 0 \\
0 & 0 & 0 & M_{\frac{d}{2}}
\end{pmatrix}
\ee
of R-charge 0, where  $M'_j$ is the bimodule map
\be
R\tc R\tc R \longrightarrow R\tc R \, , \quad 1\tc x^m \tc 1 \longmapsto \eta^{-jm} \cdot x^m\tc 1 \, .
\ee
Again, one easily computes that
\be
F_l\circ(\id\otimes F_r)\circ\alpha = (-1)^{(d-2)/2} \cdot F_r\circ(F_l\otimes\id)
\ee
holds already on the DG level, and we find the same fusing matrix $(-1)^{(d-2)/2}$ as in~\eqref{pm1}.

\bigskip

\nxt {\bf $\boldsymbol{P_{\{1,2\}}\longrightarrow (P_{\{0,1\}}\otimes P_{\{0,1\}})\otimes P_{\{0,1\}}}$:}

\medskip\noindent
Finally, we now treat the last example of subsection~\ref{subsec:CFTcalcs} involving a $(2\times 2)$-fusing matrix. In contrast to the previous simple examples it will not be sufficient to work in the DG category $\DGR(x^d)$, but we will find that the analogue of equations~\eqref{2x2eq} only holds in $\MFR(x^d)$. Also, it turns out that the calculations here are easier in the dual, i.\,e.~one-to-many, direction. As we recalled in the introduction, this will produce the same fusing matrix entries.

A possible choice of zero-charge morphisms corresponding to those which are dual to the six defect fields in~\eqref{2x2eq} is
\begin{subequations}\label{bars}
\begin{align}
A_1 & = A^{\{0,1\} \{0,1\}}_{\{1\}} =
\begin{pmatrix}
[a-(\eta+1)x+\eta b]\hat~\circ J & 0 \\
(A_1)_{21} & 0 \\
0 & J \\
0 & -\eta J
\end{pmatrix} , \\
A_2 & = A^{\{0,1\} \{0,1\}}_{\{0,1,2\}} =
\begin{pmatrix}
J & 0 \\
(A_2)_{21} & 0 \\
0 & [a+(\eta+1)x-(\eta^2+\eta+1)b]\hat~\circ J \\
0 & [(\eta^2+\eta+1)a-(\eta^2+\eta)x-\eta^2b]\hat~\circ J
\end{pmatrix} , \\
A_3 & = A^{\{0,1\} \{1\}}_{\{1,2\}} =
\begin{pmatrix}
J & 0 \\
(A_3)_{21} & 0 \\
0 & J \\
0 & [(\eta+1)a-\eta x-\eta^2 b]\hat~\circ J
\end{pmatrix} , \\
A_4 & = A^{\{0,1\} \{0,1\}}_{\{1,2\}} =
\begin{pmatrix}
(A_4)_{11} & 0 \\
(A_4)_{21} & 0 \\
0 & (A_4)_{32} \\
0 & -J
\end{pmatrix} , \\
A_5 & = A^{\{1\} \{0,1\}}_{\{1,2\}} =
\begin{pmatrix}
J & 0 \\
(A_5)_{21} & 0 \\
0 & [a + \eta x - (\eta^2+\eta)b]\hat~\circ J \\
0 & \eta^2 J
\end{pmatrix} , \\
A_6 & = A^{\{0,1,2\} \{0,1\}}_{\{1,2\}} =
\begin{pmatrix}
[a - (\eta^2+\eta+1) x + (\eta^2+\eta)b]\hat~\circ J & 0 \\
(A_6)_{21} & 0 \\
0 & J \\
0 & (A_6)_{42}
\end{pmatrix}
\end{align}
\end{subequations}
where the map $J:R\tc R\rightarrow R\tc R\tc R$ is given by $r\tc s\mapsto r\tc 1\tc s$ and we have
\begin{align*}
(A_4)_{11} & = [(\eta^{-1}+\eta^{-2})a-(1+\eta^{-1}+\eta^{-2})x-b]\hat~\circ J \vphantom{\frac{1+\eta^{-1}}{\eta+1}}\\
(A_4)_{32} & = [(\eta^{-1}+\eta^{-2})a+\frac{1+\eta^{-1}}{\eta+1}x-(\eta+1+\eta^{-1})b]\hat~\circ J \ , \\
(A_6)_{42} & = [-(\eta^3+\eta^2+\eta)a + \eta^3 x + (\eta^4+\eta^3)b]\hat~\circ J \, , \vphantom{\frac{1+\eta^{-1}}{\eta+1}}\\
(A^{IJ}_K)_{21} & = \left( (A^{IJ}_K)_{42} \hat p_{\{0,\ldots,d-1\}\backslash I}(a,x) - (A^{IJ}_K)_{32} \hat p_{\{0,\ldots,d-1\}\backslash J}(x,b)\right)/\hat p_K(a,b) \vphantom{\frac{1+\eta^{-1}}{\eta+1}} \, .
\end{align*}
One may check that the entries $(A^{IJ}_K)_{21}$ are indeed polynomials in $a,x,b$. The fusing matrix $\widetilde F=
(\begin{smallmatrix}
\widetilde F_{00} & \widetilde F_{02} \\
\widetilde F_{20} & \widetilde F_{22}
\end{smallmatrix})$ is to be determined from the equations
\begin{subequations}\label{ABC}
\begin{align}
\alpha\circ(A_1\otimes\id)\circ A_5 & = \widetilde F_{00}\cdot (\id\otimes A_1)\circ A_3 + \widetilde F_{02}\cdot (\id\otimes A_2)\circ A_4 \, , \\
\alpha\circ(A_2\otimes\id)\circ A_6 & = \widetilde F_{20}\cdot (\id\otimes A_1)\circ A_3 + \widetilde F_{22}\cdot (\id\otimes A_2)\circ A_4
\end{align}
\end{subequations}
of $(8\times 2)$-matrices. It turns out that only the constraints on $\widetilde F$ coming from the entries $(1,1)$, $(5,2)$, $(6,2)$ and $(7,2)$ can be solved without descending to cohomology, the unique solution being
\be\label{F}
\widetilde F = \frac{1}{\eta+1}
\begin{pmatrix}
-\eta & \eta^2 \\
\eta^2+\eta+1 & \eta^2
\end{pmatrix},
\ee
while the whole of equations~\eqref{ABC} can only hold in $\MFR(x^d)$. Instead of constructing a general homotopy up to which~\eqref{ABC} is satisfied, we have contended ourselves with checking that~\eqref{F} solves~\eqref{ABC} in $\MFR(x^d)$ for many values of~$d$ with the help of the computer algebra system Singular~\cite{Singular}. 

We note that~\eqref{F} is not the same fusing matrix as the result~\eqref{FCFT} but that the normalisation-invariant quantity~\eqref{Fratio} indeed produces the same result as~\eqref{Fratio2}:
\be
\frac{\widetilde F_{00} \widetilde F_{22}}{\widetilde F_{02}\widetilde F_{20}} = \frac{-\eta}{\eta^2+\eta+1} = \frac{-1}{1+2\cos\frac{2\pi}{d}} \, .
\ee

\subsubsection*{Acknowledgements}

We wish to thank Ilka Brunner, Andreas Recknagel and Sebastiano Rossi for discussions. N.~C.~thanks Ray Streater and the School of Physical Sciences and Engineering of King's College London for support. The work of I.~R.~is partially supported by the  EPSRC First Grant EP/E005047/1 and the STFC Rolling Grant ST/G000395/1.

\appendix

\section{Appendix}

\subsection{Inverses for unit isomorphisms}\label{appA}

\subsubsection*{One-variable case}

To give the inverse morphisms and the homotopies we need some more notation. Let $U,U',V,V'$ be vector spaces. If $\phi(a,b) \in \text{Hom}(U,V[a,b])$ and $\phi'(a,b) \in \text{Hom}(U',V'[a,b])$, we define the map
\be
\phi(a,x) \tc \one_R \tc \phi'(x,b) : U \tc R \tc U' \longrightarrow (V \tc R \tc V')[a,b]
\ee
to be
\be
u \tc r \tc u' \longmapsto \sum_{k,l,m,n} \phi_{kl}(u) \tc x^{l+m} r \tc \phi'_{mn}(u') \, a^k b^n \, .
\ee
That is, the formal variable $x$ acts by multiplication on the middle factor $R$. With this definition we have
\be\label{eq:hat-and-tensor}
  \hat\phi \tr \hat\phi' = [\phi(a,x) \tc \one_R \tc \phi'(x,b)]\hat~\, .
\ee
More generally we set, for $\alpha,\beta,\gamma,\delta$ each standing for one of the formal variables $a,x,b$, 
\be
\phi(\alpha,\beta) \tc \one_R \tc \phi'(\gamma,\delta) = \sum_{k,l,m} \xi_{k,l,m} \circ( \one_U \tc (x^l \cdot ~) \tc \one_{U'}) \, a^k b^m \, ,
\ee
where $\xi_{k,l,m}$ are defined by $\phi(\alpha,\beta) \tc \one_R \tc \phi'(\gamma,\delta) = \sum_{k,l,m} \xi_{k,l,m} a^k x^l b^m$. Finally, we define the maps
\begin{align}
1 \tc \one_U  :  U & \longrightarrow R \tc U \, ,
& \one_U \tc 1  :  U & \longrightarrow U \tc  R  \, , \nonumber \\
u & \longmapsto 1 \tc u \, ,
& u &  \longmapsto  u \tc 1 \, .
\end{align}
By slight abuse of notation we will abbreviate, for $\alpha,\beta \in \{a,x,b\}$, 
\begin{subequations}\label{eq:1xphi-abbrev}
\begin{align}
(1 \tc \phi(\alpha,\beta)) \equiv  (\one_R \tc \phi(\alpha,\beta)) \circ (1 \tc \one_U) \, ,\\
(\phi(\alpha,\beta)\tc 1) \equiv  (\phi(\alpha,\beta)\tc \one_R ) \circ (\one_U \tc 1)\, .
\end{align}
\end{subequations}
Note that then also 
\be\label{eq:1xphi-other}
1 \tc \phi(a,b) = (1 \tc \one_V) \circ \phi(a,b) \, , \quad
\phi(a,b) \tc 1 = (\one_V \tc 1) \circ \phi(a,b)\, .
\ee 
However, the correponding equalities
for $1 \tc \phi(x,b)$ or $\phi(a,x) \tc 1$ do not hold (or even make sense).

\medskip

It is enough to show that $\lambda_D$ and $\rho_D$ are invertible for $D = (D_0,D_1,d_0,d_1)$ where $D_0 = R \tc \check D_0 \tc R$ and $D_1 = R \tc \check D_1 \tc R$ for some vector spaces $\check D_0$, $\check D_1$. The general statement follows since every free $R$-bimodule is isomorphic to one of the form $R \tc U \tc R$ and since $\lambda_D$ and $\rho_D$ are natural in $D$.
We will show that the inverse of  $\lambda_D$ is given by
\be
  \lambda_D^{-1} = 
  \begin{pmatrix}
    [ 1 \tc  \one_{\check D_0} ]\hat~ & 0 \\
    [ \tfrac{1 \tc  \check d_0(a,b)- 1 \tc \check d_0(x,b)}{a-x}]\hat~ & 0 \\
    0 & [ \tfrac{1 \tc  \check d_1(a,b)-1 \tc \check d_1(x,b)}{a-x}]\hat~  \\
    0 & [ 1 \tc  \one_{\check D_1} ]\hat~  
  \end{pmatrix}
: D \longrightarrow I\otimes D \, .
\ee
First we need to check that $\lambda_D^{-1}$ is a morphism in $\text{MF}_{\text{bi}}(W)$, i.\,e.
\be\label{lambdainversmorph}
(I\otimes D)\circ \lambda_D^{-1} = \lambda_D^{-1}\circ D \, .
\ee
 Let us denote the left-hand side of this equation by $\mathcal L=(\begin{smallmatrix}0&\mathcal L_1\\ \mathcal L_0&0\end{smallmatrix})$ and the right-hand side by $\mathcal R=(\begin{smallmatrix}0&\mathcal R_1\\ \mathcal R_0&0\end{smallmatrix})$.  Using the composition rule below \eqref{eq:compose-ab-polys}, as well as \eqref{eq:hat-and-tensor}, \eqref{eq:1xphi-abbrev} and \eqref{eq:1xphi-other} we find
\begin{align}
(\mathcal L_0)_2 &= 
(\one_{I_0}\otimes_R d_0)\circ[1\tc \one_{\check D_0}]\hat~ + [1\tc \check d_0(a,b) - 1 \tc
\check d_0(x,b)]\hat~ \nonumber \\
&= [1\tc \check d_0(a,b)]\hat~ 
= [(1\tc \one_{\check D_1}) \circ \check d_0(a,b)]\hat~ 
= (\mathcal R_0)_2\, .
\end{align}
Similarly one gets $(\mathcal L_1)_{1}=(\mathcal R_1)_{1}$. 
Next we compute
\begin{align*}
(\mathcal L_0)_{1} - (\mathcal R_0)_{1} & = (\iota_0\otimes_R\one_{D_0})\circ[1\tc \one_{\check D_0}]\hat~ - [\tfrac{1 \tc  \check d_1(a,b)- \check d_1(x,b)}{a-x} \circ \check d_0(a,b)]\hat~ \\
& \qquad - [(\one_R\tc  \check d_1(x,b))\circ \tfrac{1 \tc  \check d_0(a,b)- \check d_0(x,b)}{a-x}]\hat~ \\
& = (\iota_0\otimes_R\one_{D_0})\circ[1\tc \one_{\check D_0}]\hat~ - 
[\tfrac{1\tc W(a) - 1 \tc  W(x)}{a-x}]\hat~ 
\end{align*}
where we used \eqref{eq:hat-and-tensor}
in the first step. If we now expand $\frac{W(a)-W(b)}{a-b}=\sum_{m,n}V_{mn}a^mb^n$ we find that the last two terms in the above computation cancel each other:
\begin{align*}
((\iota_0\otimes_R\one_{D_0})\circ[1\tc \one_{\check D_0}]\hat~)(r\tc  v\tc  s)
& = \sum_{m,n} V_{mn} rx^m\tc  x^n\tc v \tc  s \\
& = [\tfrac{1\tc W(a) - 1 \tc  W(x)}{a-x}]\hat~(r\tc  v\tc  s) \, .
\end{align*}
Similarly one shows that $(\mathcal L_1)_{2}=(\mathcal R_1)_{2}$. This means that the condition~\eqref{lambdainversmorph} is satisfied and $\lambda_D^{-1}$ is indeed a morphism.

\medskip

It remains to be seen that $\lambda_D^{-1}$ is inverse to $\lambda_D$. Because of $(\mu \otimes_R \one_{D_i}) \circ [1 \tc  \one_{\check D_i} ]\hat~ = \one_{D_i}$ it immediately follows that $\lambda_D \circ \lambda_D^{-1} = \one_D$. Conversely, however, $[1 \tc  \one_{\check D_0} ]\hat~ \circ (\mu \otimes_R \one_{D_0})$ takes $r \tc  s \tc  v \tc  t \in R \tc  R \tc  D_0 \tc  R$ to $rs \tc  1 \tc  v \tc  t$, and we are lead to introduce a map $r_{xa} \in \text{Hom}(R,R[a,b])$ given by $r_{xa}(x^n) = a^n$ (independent of $b$). This can be written as a power series
\be
  r_{xa} = \sum_{n=0}^\infty a^n \delta_{x^n}
  \quad \text{where} \quad
  \delta_{x^n} : R \longrightarrow R \, , \quad
  \delta_{x^n}(x^m) = \delta_{m,n} \, .
\ee
The infinite sum does not cause a problem because only finitely many terms are non-zero for a given element of $R$. We can write
\be
  [1 \tc  \one_{\check D_0} ]\hat~ \circ (\mu \otimes_R \one_{D_0})
  = [r_{xa} \tc  \one_{\check D_0}]\hat~ \, .
\ee
Next we note that the map $\frac{r_{xa}-1}{a-x}$ from $R$ to $R[a,b]$ is well-defined since
$(r_{xa}-1)(x^n) = a^n-x^n$, and
\be\label{useful}
\frac{r_{xa}-1}{a-x}=\sum_{n=1}^\infty\frac{a^n-x^n}{a-x}\delta_{x^n} = \sum_{n=1}^\infty\sum_{i=0}^{n-1} a^ix^{n-i-1}\delta_{x^n} \, .
\ee

\begin{lemma}\label{lem:rxa-prop}
We have
\begin{enumerate}
\item $\big[\tfrac{r_{xa}-1}{a-x} \tc \one_U \big]\hat~ \circ \big[ (a-x) \tc \one_U \big]\hat~ 
= - \big[ \one_R \tc \one_U\big]\hat~$,
\item $\big[ (a-x) \tc \one_U \big]\hat~ \circ \big[\tfrac{r_{xa}-1}{a-x} \tc \one_U \big]\hat~ 
= [1 \tc \one_U]\hat~ \circ (\mu \tr \one_{R \tc U \tc R}) - \big[ \one_R \tc \one_U\big]\hat~$,
\item for $\phi \in \text{Hom}(U,V[a,b])$,
\begin{align*}
&\big[\tfrac{r_{xa}-1}{a-x} \tc \one_U \big]\hat~ \circ \big[ \one_R \tc \phi(x,b) \big]\hat~
- \big[ \one_R \tc \phi(x,b) \big]\hat~ \circ \big[\tfrac{r_{xa}-1}{a-x} \tc \one_U \big]\hat~
\\
&= \big[ \tfrac{1 \tc \phi(a,b) - 1 \tc \phi(x,b)}{a-x} \big]\hat~ \circ (\mu \tr \one_{R \tc U \tc R}) \, .
\end{align*}
\end{enumerate}
\end{lemma}

\begin{proof}
To see (i)--(iii), apply the left hand side to $r \tc x^n \tc u \tc s$. For (i) this gives
\begin{align*}
& [\tfrac{r_{xa}-1}{a-x}\tc \one_U]\hat~((rx\tc   x^n - r\tc  x^{n+1})\tc  u\tc  s) \\
& =\,  \sum_{i=0}^{n-1}rx^{i+1}\tc  x^{n-i-1}\tc  u\tc  s - \sum_{i=0}^nrx^i\tc  x^{n-i}\tc  u\tc  s \\
& =\,  -r\tc  x^n\tc  u \tc  s \, ,
\end{align*}
for (ii) one finds
\begin{align*}
& \sum_{i=0}^{n-1} rx^{i+1}\tc  x^{n-i-1}\tc  u\tc  s - \sum_{i=0}^{n-1}rx^i\tc  x^{n-i}\tc  u\tc  s \\
&=\, ([1\tc \one_{\check D_0}]\hat~\circ(\mu\otimes_R\one_{D_0}) - \one_{I_0\otimes_R D_0})(r\tc  x^n\tc  u\tc  s) \ ,
\end{align*}
while for (iii) one has
\begin{align*}
& \sum_{k,l}\Big( \sum_{i=0}^{n+k-1}rx^i\tc x^{n+k-i-1}
- \sum_{i=0}^{n-1} rx^i\tc  x^{n+k-i-1} \Big) \tc  \phi_{kl}(u)\otimes x^l s 
\\
& =\,  \sum_{k,l}\sum_{i=0}^{k-1} rx^{n+i}\tc  x^{k-i-1}\tc  \phi_{kl}(u)\tc  x^l s \\ 
& = \, \big[ \tfrac{1 \tc \phi(a,b) - 1 \tc \phi(x,b)}{a-x} \big]\hat~( rx^n \tc u \tc s) \ .
\end{align*}
\end{proof}

We define
\be
  \psi^l_D = \begin{pmatrix}
  0 & 0 & 0 & 0 \\
  0 & 0 & 0 & [ \tfrac{ r_{xa}-1 }{ a-x } \tc  \one_{\check D_1} ]\hat~ \\
  [ \tfrac{ r_{xa}-1 }{ a-x } \tc  \one_{\check D_0} ]\hat~ & 0 & 0 & 0 \\
  0 & 0 & 0 & 0
  \end{pmatrix} 
\ee
and claim that this is a homotopy between $\lambda_D^{-1}\circ\lambda_D$ and $\id_{I\otimes D}$, i.\,e.
\be\label{homotcond}
\lambda_D^{-1}\circ\lambda_D - \id_{I\otimes D} = (I\otimes D)\circ\psi^l_D + \psi^l_D\circ (I\otimes D) \, ,
\ee
from which it follows that $\lambda_D$ really is an isomorphism in $\text{MF}_{\text{bi}}(W)$. 
To prove the claim, we denote the left-hand side of~\eqref{homotcond} by $\mathscr L=(\begin{smallmatrix}\mathscr L_0&0\\0& \mathscr L_1\end{smallmatrix})$ and the right-hand side by $\mathscr R=(\begin{smallmatrix}\mathscr R_0&0\\0& \mathscr R_1\end{smallmatrix})$. 
The equality $\mathscr{L}=\mathscr{R}$ follows in a straightforward way from lemma \ref{lem:rxa-prop}. For example
\begin{align*}
(\mathscr R_0)_{11}
&= (\iota_1\otimes_R\one_{D_0})\circ[\tfrac{r_{xa}-1}{a-x}\tc \one_{\check D_0}]\hat~
=  [1 \tc \one_U]\hat~ \circ (\mu \tr \one) - \one
= (\mathscr L_0)_{11} \, ,
\\
(\mathscr R_0)_{12} &= 0 = (\mathscr L_0)_{12} \, ,
\\
(\mathscr R_0)_{22}
&=  [\tfrac{r_{xa}-1}{a-x}\tc \one_{\check D_1}]\hat~\circ(\iota_1\otimes_R\one_{D_1})
= - \one = (\mathscr L_0)_{22} \, ,
\\
(\mathscr R_0)_{21} 
&= 
\big[\tfrac{r_{xa}-1}{a-x} \tc \one_{\check D_1} \big]\hat~ \circ \big[ \one_R \tc \check d_0(x,b) \big]\hat~
- \big[ \one_R \tc \check d_0(x,b) \big]\hat~ \circ \big[\tfrac{r_{xa}-1}{a-x} \tc \one_{\check D_0} \big]\hat~
\\
&= \big[ \tfrac{1 \tc \check d_0(a,b) - 1 \tc \check d_0(x,b)}{a-x} \big]\hat~ \circ (\mu \tr \one_{D_0}) 
= (\mathscr L_0)_{21} \, ,
\end{align*}
and similarly for the entries of $\mathscr R_1$ and $\mathscr L_1$.

The inverse of $\rho_D$ is given by
\be
  \rho_D^{-1} = 
  \begin{pmatrix}
    [ \one_{\check D_0} \tc  1 ]\hat~ & 0 \\
    [ \tfrac{\check d_0(a,x)\tc 1 - \check d_0(a,b)\tc 1}{x-b}]\hat~ & 0 \\
    0 & [ \one_{\check D_1} \tc 1 ]\hat~ \\
    0 & [ \tfrac{\check d_1(a,b)\tc 1-\check d_1(a,x)\tc 1}{x-b}]\hat~
  \end{pmatrix}
: D \longrightarrow D\otimes I \, ,
\ee
and the relevant homotopy is
\be
  \psi^r_D = \begin{pmatrix}
  0 & 0 & 0 & 0 \\
  0 & 0 & [ \one_{\check D_1} \tc \tfrac{ r_{xb}-1 }{ b-x } ]\hat~ & 0 \\
  0 & 0 & 0 & 0 \\
  - [ \one_{\check D_0} \tc \tfrac{ r_{xb}-1 }{ b-x } ]\hat~ & 0 & 0 & 0
  \end{pmatrix} .
\ee

From the above it also immediately follows that $\lambda_I=\rho_I$ as we have
\be
\rho_I \circ \lambda_I^{-1} = 
\begin{pmatrix}
(\one_{I_0} \otimes_R \mu) \circ  [ 1 \tc  \one_{\check I_0} ]\hat~ & 0 \\ 0 & (\one_{I_1} \otimes_R \mu) \circ  [ \tfrac{1 \tc  \check \iota_1(a,b)-1 \tc \check \iota_1(x,b)}{a-x}]\hat~ \end{pmatrix} ,
\ee
and it is straightforward to check that the right-hand side acts as the identity. Thus we have $\rho_I \circ \lambda_I^{-1} = \id_I$, and hence $\rho_I = \lambda_I$.

\subsubsection*{Many-variable case}

To check that the left and right unit maps~\eqref{lambdarho-gen} are indeed morphisms in $\DG(W)$, that they are natural and that they satisfy the triangle axiom is straightforward. Hence to prove theorem~\ref{thm:MFBmany} we only need to show that $\lambda_D$ and $\rho_D$ are isomorphisms in $\MF(W)$. Instead of constructing explicit homotopies as in the one-variable case, in the following we will argue more abstractly and compactly. The price to pay is that the argument only works in $\MF(W)_\text{f}$.

Let the $\C$-vector space $H(D)$ be the cohomology of a matrix (bi-)factorisation $D$ as defined in~\cite[sec.\,3]{Khovanov:2004}. $H(D)$ is finite-dimensional iff $D \in \MF(W)_\text{f}$ \cite[cor.\,5]{Khovanov:2004}.

Two morphisms $F$ and $G$ in $\text{Hom}_{\MF(W)}(D,D)$ induce linear maps $H(F)$ and $H(G)$ from $H(D)$ to itself. If we additionally have $F\circ G=\id_D$, then we also have $H(F)\circ H(G)=\id_{H(D)}$. If $D \in \MF(W)_\text{f}$ then 
$H(G)$ must be an isomorphism of vector spaces. But then according to~\cite[prop.\,8]{Khovanov:2004} also $G$ is an isomorphism in $\MF(W)$, and because of $F\circ G=\id_D$ we find $F=G^{-1}$. 

Let us now apply this reasoning to the case at hand. The inverses $\lambda_D^{-1}:D\rightarrow I\otimes D$ and $\rho_D^{-1}:D\rightarrow D\otimes I$ are most easily presented with the help of exterior algebras. However, we need not introduce this language here since for our purposes it is sufficient to state the result that these morphisms are of the form
$$
\lambda_D^{-1}=
\begin{pmatrix}
[1\tc\one_{D_0}]\hat~ & 0 \\
* & 0 \\
\vdots & \vdots \\
* & 0 \\
\hline
* & 0 \\
\vdots & \vdots \\
* & 0 \\
\hline
0 & * \\
\vdots & \vdots \\
0 & * \\
\hline
0 & [1\tc\one_{D_1}]\hat~ \\
0 & * \\
\vdots & \vdots \\
0 & * \\
\end{pmatrix}, \quad
\rho_D^{-1}=
\begin{pmatrix}
[1\tc\one_{D_0}]\hat~ & 0 \\
* & 0 \\
\vdots & \vdots \\
* & 0 \\
\hline
* & 0 \\
\vdots & \vdots \\
* & 0 \\
\hline
0 & [1\tc\one_{D_1}]\hat~ \\
0 & * \\
\vdots & \vdots \\
0 & * \\
\hline
0 & * \\
\vdots & \vdots \\
0 & * \\
\end{pmatrix}.
$$
One easily checks that $\lambda_D\circ\lambda_D^{-1}=\id_D$ and $\rho_D\circ\rho_D^{-1}=\id_D$.

In~\cite[prop.\,23]{Khovanov:2004} and \cite[sec.\,5]{br0707.0922} the existence of isomorphisms $\varphi_D^l:I\otimes D\rightarrow D$ and $\varphi_D^r:D\otimes I\rightarrow D$ was shown. Thus, if we set $F=\lambda_D\circ(\varphi_D^l)^{-1}$ and $G=\varphi_D^l\circ\lambda_D^{-1}$ (or $F=\rho_D\circ(\varphi_D^r)^{-1}$ and $G=\varphi_D^r\circ\rho_D^{-1}$) we find by the above argument that $\lambda_D$ and $\rho_D$ are indeed isomorphisms.

\subsection[Some categories for $\mathcal N=2$ minimal models]{Some categories for $\boldsymbol{\mathcal N=2}$ minimal models}\label{app:cft}

In this appendix we collect the ingredients necessary to compute the fusing matrices for the category $\mathcal{D}_d^{\mathcal N=2}$ describing the defects of $\mathcal N=2$ minimal models and sketch the computation of these fusing matrices. We do not provide all details, but we have tried to include sufficient references to places in the literature where more details can be found.

\subsubsection*{Fusing and braiding matrices}

\newcommand\Hom{\mathrm{Hom}}

Let $\mathcal{C}$ be a braided fusion category over $\C$
(see e.\,g.~\cite{baki}), that is, a
$\C$-linear
semi-simple
rigid
braided
monoidal category
with finitely many isomorphism classes of simple objects,
finite-dimensional morphism spaces,
and such that the unit object ${\bf 1}$ of $\mathcal{C}$ is simple.
Denote the monoidal structure by $(\otimes,\alpha,{\bf 1},\lambda,\rho)$ as in definition~\ref{def:monoidal} and the braiding isomorphisms by $c_{U,V} : U \otimes V \rightarrow V \otimes U$. The categories presented below are also balanced, and we denote the twist isomorphisms by $\theta_U : U \rightarrow U$. (In fact, these categories are even modular, but we will not use this.)

Select a set $\{ U_i \,|\, i \in \mathcal{I} \}$ of representatives of the isomorphism classes of simple objects in $\mathcal{C}$ such that $U_0 = {\bf 1}$. For notational simplicity we will assume that $\Hom(U_i \otimes U_j,U_k)$ is either zero- or one-dimensional. This is the case for the categories we will consider below. For every triple $i,j,k \in \mathcal{I}$ such that $\Hom(U_i \otimes U_j,U_k) \neq 0$, pick non-zero vectors
\be \label{eq:lam-in-UixUj->Uk}
 \lambda_{ij}^k \in \Hom(U_i \otimes U_j, U_k) \, , \quad
 \bar\lambda^{ij}_k \in \Hom(U_k,U_i \otimes U_j) \, ,
\ee
such that $\lambda_{ij}^k \circ \bar\lambda^{ij}_k = \id_{U_k}$. For $k \neq l$ we have $\Hom(U_k,U_l) = 0$ and so automatically $\lambda_{ij}^k \circ \bar\lambda^{ij}_l = 0$. If either $i$ or $j$ are zero, we impose the normalisation conditions
\be
 \lambda_{0i}^i = \lambda_{U_i} : {\bf 1} \otimes U_i \longrightarrow U_i \, ,
\quad
  \lambda_{i0}^i = \rho_{U_i} : U_i \otimes {\bf 1} \longrightarrow U_i \, .
\ee

\nxt {\bf Fusing matrices:} The fusing matrices implement a basis transformation between two sets of basis vectors $(U_i \otimes U_j) \otimes U_k \rightarrow U_l$ constructed from the $\lambda_{ij}^k$,
\be\label{eq:F-def-via-lambda}
 \lambda_{ip}^l \circ (\id_{U_i} \otimes \lambda_{jk}^p) \circ \alpha_{U_i,U_j,U_k}
 = \sum_{q \in \mathcal{I}} F^{(ijk)l}_{pq} \cdot
 \lambda_{qk}^l \circ (\lambda_{ij}^q \otimes \id_{U_k}) \, .
\ee
Similarly, one can use the $\bar\lambda^{ij}_k$ to construct two bases of
$U_l \rightarrow U_i \otimes (U_j \otimes U_k)$. This gives
\be\label{eq:F-def-via-lambdabar}
 \alpha_{U_i,U_j,U_k}
 \circ
 (\bar\lambda^{ij}_q \otimes \id_{U_k})
 \circ
 \bar\lambda^{qk}_l
 = \sum_{p \in \mathcal{I}} F^{(ijk)l}_{pq} \cdot
 (\id_{U_i} \otimes \bar\lambda^{jk}_p)
 \circ
 \bar\lambda^{ip}_l
\ee
for the same numbers $F^{(ijk)l}_{pq}$ as in \eqref{eq:F-def-via-lambda}.

\medskip
\nxt {\bf Braiding matrices:} The braiding matrices $R^{(ij)k} \in \C^\times$ encode the action of the braiding isomorphisms on simple objects. There are the following two equivalent expressions defining the $R^{(ij)k}$:
\be
 \lambda_{ji}^k \circ c_{U_i,U_j} = R^{(ij)k} \cdot \lambda_{ij}^k \, , \quad
 c_{U_i,U_j} \circ \bar\lambda^{ij}_k = R^{(ij)k} \cdot \bar\lambda^{ji}_k \, .
\ee

\nxt {\bf Twist eigenvalues:} The twist eigenvalues are simply the constants $\theta_k \in \C^\times$ determined by $\theta_{U_k} = \theta_k \cdot \id_{U_k}$. They are well-defined because $U_k$ is simple, which under the present assumptions implies that it is also absolutely simple, $\Hom(U_k,U_k) = \C \cdot \id_{U_k}$.

\subsubsection*{Rational free boson}

Here we describe the category $\mathcal{C}^{\mathfrak u(1)}_{2N}$. It arises as the category of representations of the rational vertex operator algebra obtained when extending the Heisenberg vertex operator algebra (the $\widehat{\mathfrak{u}}(1)$-current algebra) by two fields of weight $N$. Its fusing matrices were computed in \cite{Brunner:2000wx}, here we use the conventions of \cite[app.\,B]{Fuchs:2007tx}.

The index set for the simple objects $U_k$ is $\mathcal{I} = \{0,\dots,2N{-}1 \}$. The fusion rules are $U_m\otimes U_n \,{\cong}\, U_{[m+n]}$, where $m,n \in \mathcal{I}$ and $[m+n] = m + n\ (\text{mod}\,2N) \in \mathcal{I}$. The twist eigenvalues, braiding matrices and fusing matrices are given by
\be
 \label{eq:u(1)-th-F-R}
 \theta_k = \E^{- \pi \I k^2 / (2N) } \, , \quad
 R^{(kl)[k{+}l]} = \mathrm \E^{-\pi \mathrm \I k l / (2N) } \, , \quad
 F^{(rst)[r{+}s{+}t]}_{[s{+}t]\,[r{+}s]} = (-1)^{r\, \sigma(s{+}t)}\, ,
\ee
where $k,l,r,s,t \,{\in}\, \mathcal{I}$,
and $\sigma(k+l)$ is $0$ if $k+l < 2N$ and $1$ otherwise.

\subsubsection*{Affine $\boldsymbol{\mathfrak{su}(2)}$}

Here we describe the category $\mathcal{C}^{\mathfrak{su}(2)}_{k}$. It arises as the category of integrable highest weight representations of $\widehat{\mathfrak{su}}(2)_k$, or as a semi-simple quotient of the category of representations of $U_q(\mathfrak{sl}(2))$ with $q = \E^{\pi\I / (k+2)}$. Its data was given in \cite{Hou:1990gq,Kirillov:1991ec}; we use the conventions of \cite[sec.\,2.5.2]{tft1}.

The index set for the simple objects $U_i$ is $\mathcal{I} = \{0,\dots,k \}$. The fusion rules are
\be
 U_i \otimes U_j \cong \quad\sideset{}{^{{}_{(+2)}}} \bigoplus_{l = |i-j|}^{\min(i+j,2k-i-j)} U_l \, .
\ee
where the superscript $\,(+2)\,$ means that the direct sum is carried out
in steps of two. The twist eigenvalues and braiding matrices are given by
\be
 \theta_r = \E^{-2 \pi \I \Delta_r} \, , \quad
 R^{(rs)t} = (-1)^{(r+s-t)/2} \, \E^{-\I\pi( \Delta_t-\Delta_r-\Delta_s )}
 \, ,
\ee
where $\Delta_n = n(n+2)/(4k+8)$. The fusing matrices are
\be
 \label{eq:su(2)-Fusing-mat}
 F^{(rst)u}_{pq} = \left\{ \begin{array}{ccc}  t/2 & s/2 & p/2 \\ r/2 & u/2 & q/2 \end{array} \right\}
\ee
with
\begin{align}
 \left\{ \begin{array}{ccc} a & b & e \\[-.2em] d & c & f \end{array} \right\}
 & = (-1)^{a+b-c-d-2e} \, \sqrt{ [2e{+}1]\,[2f{+}1] } \, \Delta(a,b,e)\,\Delta(a,c,f) \nonumber
\\
& \quad \cdot \, \Delta(c,e,d) \, \Delta(d,b,f)
 ~ \sum_{z} (-1)^z \,[z{+}1]! \Big( \, [z{-}a{-}b{-}e]! \nonumber
\\  
& \quad \cdot
  \, [z{-}a{-}c{-}f]! \, [z{-}b{-}d{-}f]! \, [z{-}d{-}c{-}e]! \nonumber
\\  
& \quad \cdot \,
 [a{+}b{+}c{+}d{-}z]! \, [a{+}d{+}e{+}f{-}z]! \, [b{+}c{+}e{+}f{-}z]!
 \, \Big)^{\!-1}
 \end{align}
and
 \be  \Delta(a,b,c) :=  \sqrt{
 [-a{+}b{+}c]! \, [a{-}b{+}c]! \, [a{+}b{-}c]! \, / \,
 [a{+}b{+}c{+}1]! } \, .
\ee
The symbols $[n]$ and $[n]!$ stand for $q$-numbers and $q$-factorials, respectively, i.\,e.
 \be
 [n] = \frac{ \sin\big( {\textstyle\frac{\pi n}{k+2}} \big) }
 { \sin\big( {\textstyle\frac\pi{k+2}} \big) }\, , \qquad
 [n]! = \prod_{m=1}^n [m] \, , \qquad [0]! = 1 \,.  \ee
The range of the $z$-summation is such that the arguments are non-negative,
i.\,e.~$z$ runs over all integers (in steps of 1) from
$\max(a{+}b{+}e,a{+}c{+}f,b{+}d{+}f,c{+}d{+}e)$ to
$\min(a{+}b{+}c{+}d,a{+}d{+}e{+}f,b{+}c{+}e{+}f)$.

\subsubsection*{$\boldsymbol{\mathcal N=2}$ minimal models}

The bosonic part $(\sVir_d)_\text{bos}$ of the $\mathcal N=2$ super-Virasoro vertex operator algebra with central charge $c_d = 3-\frac{6}{d}$ can be obtained from the coset construction for $\big(\widehat{\mathfrak{su}}(2)_{d-2} \oplus \widehat{\mathfrak{u}}(1)_4 \big) / \widehat{\mathfrak{u}}(1)_{2d}$ \cite{Kazama:1988qp}. Using the results of \cite{Frohlich:2003hm,Frohlich:2003hg} to obtain the representation category of the coset theory from those of the component theories, one finds that the category of $(\sVir_d)_\text{bos}$-representations can be described as
\be \label{eq:CdN=2-as-locmod}
 \mathcal{C}_d^{\mathcal N=2} = \big( \mathcal{C}^{\mathfrak{su}(2)}_{d-2} \boxtimes \bar{\mathcal{C}}^{\,\mathfrak u(1)}_{2d} \boxtimes \mathcal{C}^{\mathfrak u(1)}_{4} \big)^\text{loc}_C \, .
\ee
Here, the overline in $\bar{\mathcal{C}}^{\,\mathfrak u(1)}_{2d}$ means that we replace the braiding and twist by their inverses, and $\boxtimes$ is the product of $\C$-linear categories. That is, one takes triples $U_l \times U_m \times U_s$ of objects and tensor products (over $\C$) of morphism spaces, and completes the resulting category with respect to direct sums of objects. Let us abbreviate the simple objects as
\be
 U_{l,m,s} \equiv U_l \times U_m \times U_s \in \mathcal{C}^{\mathfrak{su}(2)}_{d-2} \boxtimes \bar{\mathcal{C}}^{\,\mathfrak u(1)}_{2d} \boxtimes \mathcal{C}^{\mathfrak u(1)}_{4} \, .
\ee
The subscript $C$ in \eqref{eq:CdN=2-as-locmod} stands for the object
\be
 C = {\bf 1} \oplus J \quad \text{with} \quad  J = U_{d-2,d,2} \, .
\ee
The object $J$ obeys $J \otimes J \cong {\bf 1}$ and $\theta_J = \id_J$. It follows from \cite[def.\,3.17 \& prop.\,3.22]{tft3} that $C$ carries a unique (up to isomorphism) structure of a commutative simple special symmetric Frobenius algebra (see \cite[sec.\,2]{Frohlich:2003hm} for definitions). Finally, the notation $( \,\cdot\, )^\text{loc}_C$ in \eqref{eq:CdN=2-as-locmod} denotes the category of local $C$-modules (see \cite[sec.\,3.4]{Frohlich:2003hm} for a definition of local modules). By \cite{Kirillov:2001ti} (see also \cite[prop.\,3.21]{Frohlich:2003hm}), this is again a braided monoidal category.

We proceed to compute the category $\mathcal{C}_d^{\mathcal N=2}$. Let $\mathrm{Ind}_C(V) = C \otimes V$ denote the $C$-module induced by $V$. The $C$-action on $C\otimes V$ is given by the multiplication of~$C$. As there is no object $U_{l,m,s}$ for which $U_{l,m,s} \otimes J \cong U_{l,m,s}$, the induced modules $\mathrm{Ind}_C(U_{l,m,s})$ are all simple, and we have
\begin{align}
& \mathrm{Ind}_C(U_{l,m,s}) \cong \mathrm{Ind}_C(U_{l',m',s'}) \nonumber
\\
& \Leftrightarrow \quad (l,m,s) = (l',m',s')
 \text{ or } (l,m,s) = (d-2-l',d+m',2+s') \, . \label{eq:induced-C-modules-iso}
 \end{align}
Every $C$-module is a submodule of an induced module \cite[lem.\,4.15]{Fuchs:2001qc}, so that we have found all simple $C$-modules. All other $C$-modules are isomorphic to direct sums of these, as the category of $C$-modules is semi-simple \cite[prop.\,5.24]{Fuchs:2001qc}. A simple $C$-module $M$ is local iff $\theta_M = \xi_M \cdot \id_M$ for some $\xi_M \in \C^\times$ \cite[cor.\,3.18]{Frohlich:2003hm}. In the present case, this means that $\mathrm{Ind}_C(U_{l,m,s})$ is local iff $\theta_{l,m,s} = \theta_{d-2-l,d+m,2+s}$. One checks that
\be
 \mathrm{Ind}_C(U_{l,m,s}) \text{ local}
 \quad \Leftrightarrow \quad l+m+s \in 2 \Z \, .
\ee
Altogether, we see that the isomorphism classes of simple objects in $\mathcal{C}_d^{\mathcal N=2}$ are labelled by $\Z_2$-orbits in the set of labels $(l,m,s)$ for which $l+m+s \in 2 \Z$, with $\Z_2$-action given by $(l,m,s) \mapsto (d{-}2{-}l,d{+}m,2{+}s)$. Let $[l,m,s]$ denote a fixed choice of representative on each orbit, i.\,e.\ if $[l,m,s] = (l',m',s')$ then also $[d{-}2{-}l,d{+}m,2{+}s] = (l',m',s')$.
We denote the simple objects of $\mathcal{C}_d^{\mathcal N=2}$ by
\be
 R_{l,m,s} := \mathrm{Ind}_C(U_{l,m,s}) \, .
\ee
Let $\mathcal{E}_d$ be the full subcategory of $\mathcal{C}^{\mathfrak{su}(2)}_{d-2} \boxtimes \bar{\mathcal{C}}^{\,\mathfrak u(1)}_{2d} \boxtimes \mathcal{C}^{\mathfrak u(1)}_{4}$ consisting of all objects that are isomorphic to direct sums of simple objects $U_{l,m,s}$ with $l+m+s \in 2\Z$. It is easy to see that $\mathcal{E}_d$ is  a monoidal subcategory. By definition of the tensor product and the braiding of $C$-modules \cite[sec.\,3.4]{Frohlich:2003hm}, the induced modules provide a braided monoidal functor
\be \label{eq:IndC-braided-tensor-fun}
 \mathrm{Ind}_C(\,\cdot\,) : \mathcal{E}_d \longrightarrow \mathcal{C}_d^{\mathcal N=2} \, .
\ee
We have already seen that this functor is essentially surjective, however by \eqref{eq:induced-C-modules-iso} it is not full. (It is, however, injective on morphisms.)

In making the connection to matrix factorisations, we are primarily interested in the simple objects $U_{l,m,0}$ and their direct sums. Let $\mathcal{E}_{d,s=0}$ be the full subcategory of $\mathcal{E}_{d}$ consisting of all objects isomorphic to direct sums of simple objects of the form $U_{l,m,0}$. This is again a monoidal subcategory, in fact we can canonically identify
\be\label{eq:E_0d-is-CxC}
 \mathcal{E}_{d,s=0} = \mathcal{C}^{\mathfrak{su}(2)}_{d-2} \boxtimes \bar{\mathcal{C}}^{\,\mathfrak u(1)}_{2d} \, .
\ee
Denote by $\mathcal{C}_{d,s=0}^{\mathcal N=2}$ the full subcategory of $\mathcal{C}_{d}^{\mathcal N=2}$ consisting of objects isomorphic to $\mathrm{Ind}_C(V)$ for some $V \in \mathcal{E}_{d,s=0}$. Because \eqref{eq:induced-C-modules-iso} never relates two modules in $\mathcal{C}_{d,s=0}^{\mathcal N=2}$, we now have a braided monoidal equivalence
\be \label{eq:Cd-N=2-for-s=0}
 \mathrm{Ind}_C(\,\cdot\,) : \mathcal{E}_{d,s=0} \stackrel{\sim}{\longrightarrow} \mathcal{C}_{d,s=0}^{\mathcal N=2}  \, .
\ee

\subsubsection*{Defects for $\boldsymbol{\mathcal N=2}$ minimal models}

Let $G$ be the object $G = R_{0,0,2} \in \mathcal{C}_d^{\mathcal N=2}$ and set $S = {\bf 1} \oplus G$. There is a unique (up to isomorphism) structure of a simple special symmetric Frobenius algebra on~$S$ \cite[def.\,3.17\,\&\,prop.\,3.22]{tft3}. This Frobenius algebra is non-commutative. The defect category $\mathcal{D}_d^{\mathcal N=2}$ is given by the category of $S$-bimodules in $\mathcal{C}_d^{\mathcal N=2}$,
\be
 \mathcal{D}_d^{\mathcal N=2} = 
  S\text{-mod-}S \, .
\ee
The category $\mathcal{D}_d^{\mathcal N=2}$ is again semi-simple (apply \cite[prop.\,5.24]{Fuchs:2001qc} to $(S \otimes S^\mathrm{op})$-modules). Given an object $V \in \mathcal{C}_d^{\mathcal N=2}$ and an algebra automorphism $\psi \in \mathrm{Aut}(S)$, denote by $\alpha^+_S(V)_\psi$ the $S$-bimodule obtained via braided induction as follows (see \cite[sec.\,5.4]{tft1} for definitions and references). The underlying object is $S \otimes V$, the left action is by multiplication, $\rho^l = m \otimes \id_V$ (we omit the coherence isomorphisms), and for the right action one uses the braiding and precomposes the action by $\psi$,
\be
 \rho^r = ( m \otimes \id_V) \circ (\id_S \otimes c_{V,S}) \circ (\id_S \otimes \id_V \otimes \psi) \, .
\ee
The group $\mathrm{Aut}(S)$ is isomorphic to $\Z_2$, its non-trivial element is the isomorphism $\omega = \id_{\bf 1} - \id_G$. We have, for $\psi \in \mathrm{Aut}(S)$,
\be \label{eq:alpha-S-relates-V-GV}
 \alpha^+_S(V)_\psi \cong \alpha^+_S(G \otimes V)_{\omega \circ \psi} \, .
\ee
This follows from \cite[def.\,5.2\,\&\,prop.\,5.10]{Frohlich:2006ch} and the fact that $\theta_G = - \id_G$. If $V$ is simple, then $\alpha^+_S(V)_\psi$ is simple as an $S$-bimodule and $\alpha^+_S(V)_\psi \cong \alpha^+_S(V)_\xi$ iff $\psi = \xi$, see \cite[eq.\,(5.19)\,\&\,prop.\,5.9]{Frohlich:2006ch}. In particular,
$\alpha^+_S(V)_{\id} \ncong \alpha^+_S(G \otimes V)_{\id}$, and altogether, for $V$ and $V'$ simple,
\be \label{eq:alpha-simple-iso}
 \alpha^+_S(V)_{\id} \cong \alpha^+_S(V')_{\id}
 \quad \Leftrightarrow \quad
 V \cong V' \, .
\ee
Braided induction (with trivial automorphism) provides a monoidal functor
\be \label{eq:alpha-iso-Cd-Dd}
 \alpha^+_S(\,\cdot\,)_{\id} : \mathcal{C}_d^{\mathcal N=2} \longrightarrow \mathcal{D}_d^{\mathcal N=2} \, .
\ee
Every $S$-bimodule is a subbimodule of an $\alpha^+_S(W)_\psi$ for some $W  \in \mathcal{C}_d^{\mathcal N=2}$, $\psi \in \mathrm{Aut}(S)$ \cite[prop.\,5.7]{Frohlich:2006ch}. In fact, by \eqref{eq:alpha-S-relates-V-GV}, every $S$-bimodule is a subbimodule of an $\alpha^+_S(W)_{\id}$ for some $W  \in \mathcal{C}_d^{\mathcal N=2}$, so that $\alpha^+_S(\,\cdot\,)_{\id}$ is essentially surjective. Finally, \eqref{eq:alpha-simple-iso} implies that $\alpha^+_S(\,\cdot\,)_{\id}$ is fully faithful, and hence an equivalence.

Let us write $D_{l,m,s}$ for $\alpha^+_S(R_{l,m,s})_{\id}$. Let $\mathcal{D}_{d,s=0}^{\mathcal N=2}$ denote the full subcategory of $\mathcal{D}_{d}^{\mathcal N=2}$ consisting of objects isomorphic to direct sums of simple objects of the form $D_{l,m,0}$. Combining the equivalences \eqref{eq:E_0d-is-CxC}, \eqref{eq:Cd-N=2-for-s=0} and \eqref{eq:alpha-iso-Cd-Dd} we get a monoidal equivalence
\be
  \mathcal{F} : \mathcal{C}^{\mathfrak{su}(2)}_{d-2} \boxtimes \bar{\mathcal{C}}^{\,\mathfrak u(1)}_{2d}
  \longrightarrow \mathcal{D}_{d,s=0}^{\mathcal N=2} \, .
\ee
This makes it easy to compute the fusing matrices of $\mathcal{D}_{d,s=0}^{\mathcal N=2}$, namely they are just the product of those of $\mathcal{C}^{\mathfrak{su}(2)}_{d-2}$ and $\mathcal{C}^{\mathfrak u(1)}_{2d}$ (the overline does not affect the fusing matrices). In more detail, define the non-zero morphisms
\be
 \lambda_{(i,a)(j,b)}^{(k,c)} \in
 \text{Hom}_{\mathcal{D}_d^{\mathcal N=2}}( D_{i,a,0} \otimes D_{j,b,0} , D_{k,c,0} )
 \,  , \quad
 \lambda_{(i,a)(j,b)}^{(k,c)} =
 \mathcal{F}\big( \lambda[\mathfrak{su}(2)]_{ij}^k \times \lambda[\mathfrak u(1)]_{ab}^c \big)
\ee
where $\lambda(\mathfrak{su}(2))_{ij}^k$ refers to the basis \eqref{eq:lam-in-UixUj->Uk} chosen for $\mathcal{C}^{\mathfrak{su}(2)}_{d-2}$ and $\lambda(\mathfrak u(1))_{ab}^c $ to that chosen for $\mathcal{C}^{\,\mathfrak u(1)}_{2d}$. Then
\begin{align}%
&\lambda_{(i,a)(p,e)}^{(l,d)} \circ (\id \otimes \lambda_{(j,b)(k,c)}^{(p,e)}) \circ \alpha \nonumber \\
=&\sum_{q=0}^{d-2} \sum_{f=0}^{2d-1} F[\mathfrak{su}(2)]^{(ijk)l}_{pq} \cdot F[\mathfrak u(1)]^{(abc)d}_{ef} \cdot
 \lambda_{(q,f)(k,c)}^{(l,d)} \circ (\lambda_{(i,a)(j,b)}^{(q,f)} \otimes \id) \, ,
\end{align}
where $F[\mathfrak{su}(2)]$ stands for the fusing matrix \eqref{eq:su(2)-Fusing-mat}, and $F[\mathfrak u(1)]$ for the fusing matrix \eqref{eq:u(1)-th-F-R}.

\subsubsection*{Spectra of defects fields}

For two objects $D,E \in \mathcal{D}_d^{\mathcal N=2}$, let $H^{D \rightarrow E}_\text{bos}$ be the space of bosonic defect changing fields that change the defect labelled $D$ to the defect labelled $E$. It decomposes as
\be
 \label{eq:def-spec-D->E}
 H^{D \rightarrow E}_\text{bos}
 = \bigoplus_{[l,m,s],[l',m',s']} \big(R_{[l,m,s]} \otimes \bar R_{[l',m',s']}\big)^{M_{l,m,s;l',m',s'}} \, ,
\ee
where the multiplicities are given by the dimension of a space of $S$-bimodule isomorphisms \cite[sec.\,2.2]{Frohlich:2006ch}
\be
 M_{l,m,s;l',m',s'} = \dim \Hom_{S\text{-mod-}S}(R_{l,m,s} \otimes^+ D \otimes^- R_{l',m',s'}, E)\, .
\ee
The definition of $\otimes^\pm$ is given in \cite[sec.\,2.1]{Frohlich:2006ch}. From \eqref{eq:u(1)-th-F-R} and \eqref{eq:IndC-braided-tensor-fun} it follows that the double braiding of $G$ with a simple object $R_{l,m,s} \in  \mathcal{C}_d^{\mathcal N=2}$ is given by
\be \label{eq:G-double-braid}
 c_{R_{l,m,s},G} \circ c_{G,R_{l,m,s}} = (-1)^s \cdot \id_G \otimes \id_{R_{l,m,s}}\, .
\ee
If $D = \alpha^+_S(U)_{\id}$ for some $U$, then from \cite[prop.\,5.9]{Frohlich:2006ch} together with \eqref{eq:alpha-S-relates-V-GV} and \eqref{eq:G-double-braid} we get
\be
 R_{l,m,s} \otimes^+ \alpha^+_S(U)_{\id} \otimes^- R_{l',m',s'}
 \cong
 \alpha^+_S( R_{l,m,s} \otimes U \otimes R_{l',m',s'} \otimes G^{\otimes s'} )_{\id} \, .
\ee
If now $D = \alpha^+_S(U)_{\id}$ and $E = \alpha^+_S(V)_{\id}$ for simple $U,V \in  \mathcal{C}_d^{\mathcal N=2}$, then together with \eqref{eq:alpha-simple-iso} this implies
\begin{align}
 M_{l,m,s;l',m',s'} &= \dim \Hom_{S\text{-mod}S}(
 \alpha^+_S( R_{l,m,s} \otimes U \otimes R_{l',m',s'} \otimes G^{\otimes s'} )_{\id}  \, ,
 \alpha^+_S(V)_{\id} ) \nonumber
\\
 &= \dim \Hom( R_{l,m,s} \otimes R_{l',m',-s'} , U^\vee \otimes V ) \, ,
\end{align}
where $U^\vee$ is the dual of $U$. In the special case that $U = R_{0,2n,0}$ and $V = R_{u,2n+u,0}$ we have $U^\vee \otimes V \cong R_{u,u,0}$.

The above calculation leads to a space of bulk fields (or defect fields on the identity defect) which is a direct sum of terms $R_{l,m,s} \otimes_{\C} R_{l,-m,s}$. This is not quite what we want for the application in the main text, but it is related to that by reversing the sign of the anti-holomorphic $U(1)$-current, which amounts to replacing
\be
 R_{l,m,s} \otimes_{\C} R_{l',-m',s'} \longmapsto 
 R_{l,m,s} \otimes_{\C} R_{l',m',-s'} \, .
\ee
This replacement has to be applied to \eqref{eq:def-spec-D->E} before comparing to the statements in section~\ref{sec:apps}. (Alternatively, we could have started with a more complicated simple special symmetric Frobenius algebra $S$, which directly describes the correct modular invariant.)

\end{document}